\documentclass[iop, apjl]{emulateapj}
\usepackage{threeparttable}
\usepackage{multirow}
\usepackage{times}
\usepackage{enumitem}
\usepackage{url} 
\usepackage{amsmath,bm}
\usepackage{color,graphicx}
\usepackage{appendix}

\begin{document}
\bibliographystyle{plain}
\title{On the origins of extreme velocity stars as revealed by large-scale Galactic surveys}
\author{Qing-Zheng Li\altaffilmark{1,2}}
\author{Yang Huang\altaffilmark{2,3,8}}
\author{Xiao-Bo Dong\altaffilmark{1,8}}
\author{Hua-Wei Zhang\altaffilmark{4,5}}
\author{Timothy C. Beers\altaffilmark{6}}
\author{Zhen Yuan\altaffilmark{7}}
\altaffiltext{1}{Yunnan Observatories, Chinese Academy of Sciences, Kunming, Yunnan 650011, P.\,R.\,China}
\altaffiltext{2}{School of Astronomy and Space Science, University of Chinese Academy of Sciences, Beijing 100049,  People's Republic of China}
\altaffiltext{3}{Key Laboratory of Space Astronomy and Technology, National Astronomical Observatories, Chinese Academy of Sciences, Beijing 100101, China'}
\altaffiltext{4}{Department of Astronomy, School of Physics, Peking University, Beijing 100871, P.\,R.\,China}
\altaffiltext{5}{Kavli Institute for Astronomy and Astrophysics, Peking University, Beijing 100871, P.\,R.\,China}
\altaffiltext{6}{Department of Physics and Astronomy and JINA Center for the Evolution of the Elements (JINA-CEE), University of Notre Dame, Notre Dame, IN 46556, USA}
\altaffiltext{7}{Universit{\'e} de Strasbourg, CNRS, Observatoire Astronomique de Strasbourg, UMR 7550, F-67000 Strasbourg, France}
\altaffiltext{8}{Corresponding authors: {\it \mbox{huangyang@bao.ac.cn} {\rm (YH)}}; {\it \mbox{xbdong@ynao.ac.cn} {\rm (XBD)}}}
\begin{abstract}
We assemble a large sample of 12,784\,high-velocity stars with total velocity $\it{V}_{\rm{GSR}}\ge{\rm300}\,\rm{km\,s^{-1}}$, selected from RAVE\,DR5, SDSS\,DR12, LAMOST\,DR8, APOGEE\,DR16, GALAH\,DR2, and {\it\,Gaia}\,EDR3. 
In this sample, 52\,are marginally hypervelocity star (HVS) candidates that have $\it{V}_{\rm{GSR}}$ exceeding their local escape velocities within $2\sigma$ confidence levels, 40\,of which are discovered for the first time. 
All candidates are metal-poor, late-type halo stars, significantly different from the previous identified HVSs, which are largely massive early-type stars, discovered by extreme radial velocity. 
This finding suggests that our newly identified HVS candidates are ejected by different mechanisms from the previous population. 
To investigate their origins, for 547\,extreme velocity stars with ${V}_{\rm{GSR}}\ge0.8V_{\rm{esc}}$, we reconstruct their backward-integrated trajectories in the Galactic potential. 
According to the orbital analysis, no candidates are found to be definitely ejected from the Galactic-center (GC), while 8 metal-poor extreme velocity stars are found to have a closest distance to the GC within 1kpc. 
Intriguingly, 15\,extreme velocity stars (including 2\,HVS candidates) are found to have experienced close encounters with the Sagittarius dSph, suggesting that they originated from this dSph. 
This hypothesis is supported by an analysis of the [$\alpha$/Fe]--[Fe/H] diagram. 
From a preliminary analysis of all the 547 extreme velocity stars, we propose a general picture: Star ejection from Galactic subsystems such as dwarf galaxies and globular clusters can be an important channel to produce extreme velocity stars or even HVSs, particularly the metal-poor late-type halo population.
\end{abstract}

\section{Introduction}

\begin{figure*}
\begin{center}
\includegraphics[scale=0.52,angle=0]{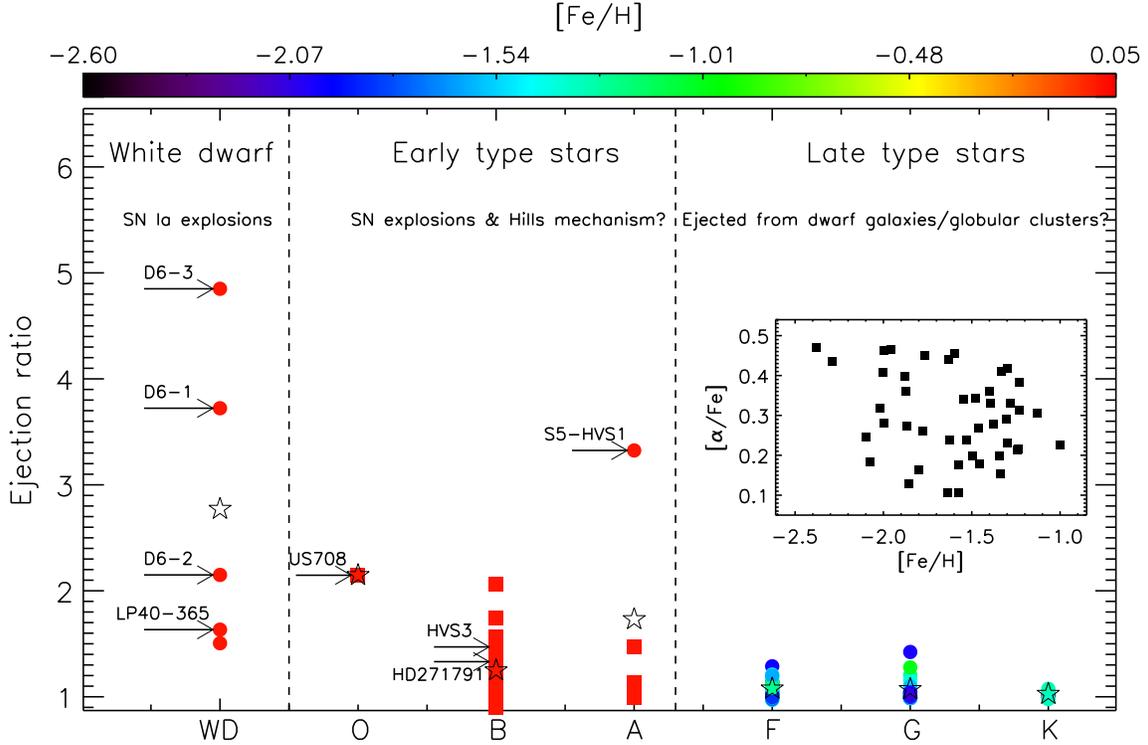}
\caption{Ejection ratios for 88 HVSs and candidates from the literature and the current work, as a function of spectral type, color-coded by metallicity. Here we adopt the escape velocity curve from \citet{Williams2017} for calculating the ejection ratio. As mentioned in Section 3.3, the 52 late-type stars found in this work are mariginally HVS candidates given their large velocity uncertainties.
	The ejection ratio is defined to be the ratio of the total or radial velocity in the Galactic rest frame to the local escape velocity of a star.  
	Note that the [Fe/H] values for white dwarf and early-type stars are assumed to be Solar ([Fe/H] = zero). The open asterisk marks the average ejection ratio of the corresponding spectral type. 
The inset shows the chemical distribution of the late-type stars discovered in the current work in the [$\alpha$/Fe]--[Fe/H] plane.}
\end{center}
\end{figure*}

The majority of stars in our Galaxy either rotate around the Galactic center (GC) with a typical velocity of 200--240\,km\,s$^{-1}$ \citep[e.g.,][]{Bovy2012,Huang2016,Eilers2019} in the disk region, 
or exhibit large random motions of 100--150\,km\,s$^{-1}$ in the halo \citep{Xue2008,Huang2016}.  
But over the past decade, spectroscopic observations from large-scale Galactic surveys 
such as the Sloan Extension for Galactic Understanding and Exploration (SEGUE, \citealt{Yanny2009}; SEGUE-2, \citealt{Rockosi2022}), the LAMOST Galactic surveys \citep{Deng2012,Zhao2012,Liu2014}, and the {\it Gaia} mission \citep{GaiaCollaboration2016}, 
have demonstrated the existence of high-velocity stars (HiVels) in our Galaxy, 
some of which are even hypervelocity stars (HVSs), with total Galactocentric velocities, $\it{V}_{\rm{GSR}}$,  exceeding their local escape speeds, $V_{\rm esc}$.
The discoveries of such rare objects provide an important tool to explore the mass distribution of the Milky Way, especially its dark component \citep[e.g.,][]{Gnedin2005,Rossi2017,Contigiani2019}.

Over the past few decades several ejection mechanisms have been proposed to explain the presence of HiVels/HVSs in the Milky Way,
as briefly summarized below.

\textbf{Black hole ejection (BHE):} 
The so-called HVSs (with velocities even greater than 1000 km\,s$^{-1}$) were first predicted from the theoretical arguments of \citet{Hills1988}, and attributed
to be the result of tidal interaction between a close stellar binary system and a supermassive black hole (SMBH) in the GC, commonly referred to as the ``Hills mechanism." 
HVSs/HiVels can also be ejected by extending the Hills mechanism with alternative assumptions: 
1) a SMBH binary or the pair of a SMBH and an intermediate-mass black hole (IMBH) in the GC \citep[e.g.,][]{Yu2003, Gualandris2009,Rasskazov2019}, 
or 2) individual IMBHs (or SMBHs) in the Galactic dwarf galaxies or globular clusters,
or even massive BH binaries in such Galactic subsystems  \citep[e.g.,][]{Boubert2016,Fragione2019}. 

\textbf{Supernova explosion (SNE) in binary systems:} Both core-collapse and thermonuclear supernova (SN) explosions in binary systems 
can disrupt the system and kick their companion stars sufficiently for them to become HiVels/HVSs 
\citep[e.g.,][]{Blaauw1961,PortegiesZwart2000,Hansen2003,Justham2009,Wang2009, Pakmor2013,Zubovas2013,Shen2018,Bauer2019,Neunteufel2020}.
Typically, the core-collapse SN explosions cannot eject stars with very high velocity \citep[no greater than 300--400\,km\,s$^{-1}$;][]{PortegiesZwart2000}.
This mechanism is mainly to account for the existence of OB runaway stars above the Galactic plane \citep{Blaauw1961}.
HVSs with velocities above 1000\,km\,s$^{-1}$ can be ejected in thermonuclear SN explosions, with progenitors of either white dwarf (WD)--WD binaries \citep[e.g.,][]{Shen2018} or binaries comprising a WD plus a non-degenerate star \citep[e.g.,][]{Han2008,Bauer2019,Wang2009,Neunteufel2020}.

\textbf{Dynamical ejection mechanism (DEM):} This was an alternative theory to explain the Galactic OB runaway stars first proposed by \citet{Poveda1967}.
In this mechanism, runaway stars were ejected as a consequence of close stellar encounters in young stellar clusters.
The typical maximum kick velocity achieved by this mechanism is around 300--400\,km\,s$^{-1}$, resulting from collisions between two close binaries \citep[e.g.,][]{Leonard1990,Leonard1991,Gvaramadze2009}.

\textbf{Tidal stripping from dwarf galaxies (TSD):} 
According to this theory, stars can be stripped with high velocity from a dwarf galaxy being tidally disrupted by the gravity field of the Milky Way (MW) during its pericentric passage \citep{Abadi2009}.  In this mechanism, a massive dwarf galaxy ($> 10^{10}$\,M$_{\odot}$) is required to eject unbound stars \citep{Piffl2011}.

Observationally, the first HVS, a B-type star with an extreme radial velocity of 709 km\,s$^{-1}$ in the Galactic rest frame, 
was discovered by \citet{Brown2005}. 
After that, over two dozen HVSs, all being early-type, 
were either serendipitous discoveries \citep{Hirsch2005,Heber2008,Koposov2020}, or resulted from dedicated follow-up surveys \citep{Brown2006,Brown2007,Brown2009,Brown2012,Brown2014,Zheng2014,Huang2017,Li2018}.
The only low-mass HVS among these early-type stars is US\,708, an O-type subdwarf compact helium star with a total velocity of around 994\,km\,s$^{-1}$ \citep[][see Fig.\,1]{Hirsch2005,Geier2015,Neunteufel2020}, which was suggested to be ejected from a SN Ia explosion \citep{Geier2015}.
The remaining HVSs are young, massive B/A-type stars, mostly identified from their extreme radial velocities only. 
Among them, HVS3 (or HE\,0437-5439) was suggested to have been ejected from the Large Magellanic Cloud (LMC) via the Hills mechanism \citep{Edelmann2005}, which was confirmed by the {\it Gaia} data \citep{Irrgang2018,Erkal2019}. HD~271791 was suggested to have been ejected from the Galactic disk, either via DEM \citep{Gvaramadze2009} or SNE \citep{Przybilla2008}.
Recently, S5-HVS1, an A-type star with a total velocity of $1755 \pm 50$\,km\,s$^{-1}$, was discovered to point unambiguously to the GC, 
based on its backward-integrated trajectory  \citep{Koposov2020}, 
providing solid evidence of the operation of the Hills mechanism in our Galaxy. We note there have been previous efforts searching for late-type HVSs \citep{Li2012,Li2015,Palladino2014}, but almost all of these are likely bound to our Galaxy, with a single exception (LAMOSTJ115209.12+120258.0), as revisited by \citet{Ziegerer2015} or \citet{Boubert2018}.
Fig.\,1 shows the ejection ratio (defined as the ratio between the total or radial velocity in the Galactic rest frame and the local escape velocity), 
as a function of spectral type, for the known HVSs and candidates (including those found in the current work).

Thanks to the \textit{Gaia} mission \citep{GaiaCollaboration2018,GaiaCollaboration2021},  
several dozens of new HVSs and candidates have been discovered, using very precise astrometric parameters \citep[e.g.,][]{Bromley2018,Du2018,Du2019,Shen2018,Irrgang2019,
Marchetti2019,Marchetti2021,Li2020}.
Among those sources, \citet{Shen2018} found three hypervelocity WDs.
These three WDs, together with LP~40-365 and three other potential high velocity WDs \citep{Vennes2017,Raddi2018,Raddi2019}, provide direct evidence for the operation of the SNE ejection mechanism in the Galaxy \citep[especially the dynamically driven double-degenerate double-detonation--D6 channel;][]{Shen2018}.
Astrometry from {\it Gaia} has also enabled more precise measurements of 3D velocities for those known massive B/A-type HVSs, which further improves our understanding of their origins \citep{Hattori2019,Kreuzer2020,Irrgang2021}.

In this paper, we present the results of a new systematic search for HiVels from the combination of astrometric data from the \textit{Gaia} early data release 3 (EDR3) and large-scale spectroscopic surveys, including  the RAVE\,DR5, SDSS\,DR12, LAMOST\,DR8, APOGEE\,DR16, and GALAH\,DR2 surveys.
A sample of 12,784 HiVels (including 6966 halo stars and 5818 disk stars) with total velocities $V_{\rm GSR} \ge 300$\,km\,s$^{-1}$ are found.
Interestingly, 52 of them are HVS candidates but with total velocities marginally exceeding the local escape velocities (see Fig.\,1).  All of these candidates are late-type metal-poor stars, 
significantly different from those HVSs in the literature (primarily massive early-type stars) 
that were mostly found by their extreme radial velocities alone (see Fig.\,1). Our newly discovered late-type HVS candidates, together with the known ones, shed additional light on their likely ejection mechanisms.

This paper is organized as follows. 
In Sections\,2, we briefly introduce the data we employ. 
In Sections\,3 and 4, we present the main results and compare them with previous studies. 
We discuss the origins of the newly discovered extreme velocity stars with $V_{\rm GSR} \ge 0.8V_{\rm esc}$ in Section\,5.
Finally, a summary is presented in Section\,6.

\begin{table}[!t]
\tablename{1: Comparisons of Radial Velocities between Large-scale Spectroscopic Surveys and Radial Velocity Standard Stars.}
\centering
\begin{threeparttable}
\begin{tabular}{rcrrr}
\toprule
\multirow{1}{*} &  Surveys  &  $\Delta v_{\rm los} (\rm{km\,s^{-1}})$  &  s.d.\,$(\rm{km\,s^{-1}})$ & $N$ \\
\cline{1-5}
\multirow{1}{*} &    GALAH  &                     $-0.31$  &                       0.52 &76 \\ 
\multirow{1}{*} &   APOGEE  &                      $0.28 $ &                       0.19  &18,080\\
\multirow{1}{*} &     RAVE  &                      $0.07$  &                       1.37  &374\\
\multirow{1}{*} &   LAMOST  &                     $-4.97$  &                       3.83 & 4975 \\
\cline{1-5}
\end{tabular}
\end{threeparttable}
\end{table}

\section{Data}
In the current work, we use data from modern large-scale Galactic spectroscopic surveys, including the RAVE \citep{Steinmetz2006}, SDSS \citep{Alam2015}, LAMOST \citep{Deng2012,Liu2014}, APOGEE \citep{Majewski2017} and GALAH \citep{DeSilva2015} surveys, as well as from the \textit{Gaia} astrometric survey \citep{GaiaCollaboration2016,GaiaCollaboration2018,GaiaCollaboration2021}.

\subsection{Spectroscopic Surveys}

\begin{figure}[!t]
\begin{center}
\includegraphics[scale=0.17,angle=0]{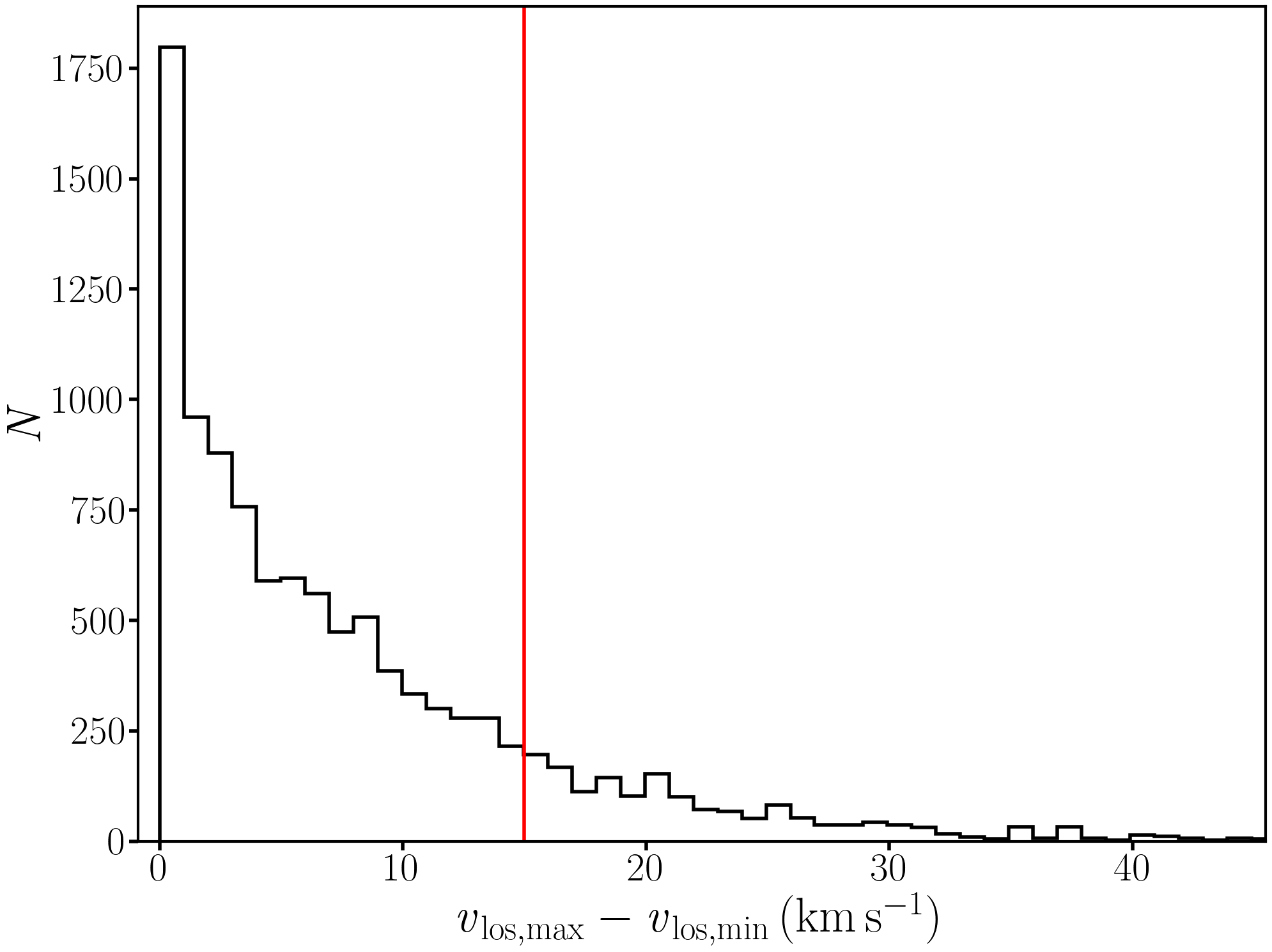}
\caption{The $\it{v}_{\rm{los}}$ variations (the maximum minus the minimum value) for our 4013 HiVel candidates with multiple observations. For reference, the red vertical line marks a difference of 15\,km\,s$^{-1}$.}
\end{center}
\end{figure}

\begin{figure*}[!t]
\begin{center}
\includegraphics[scale=0.22,angle=0]{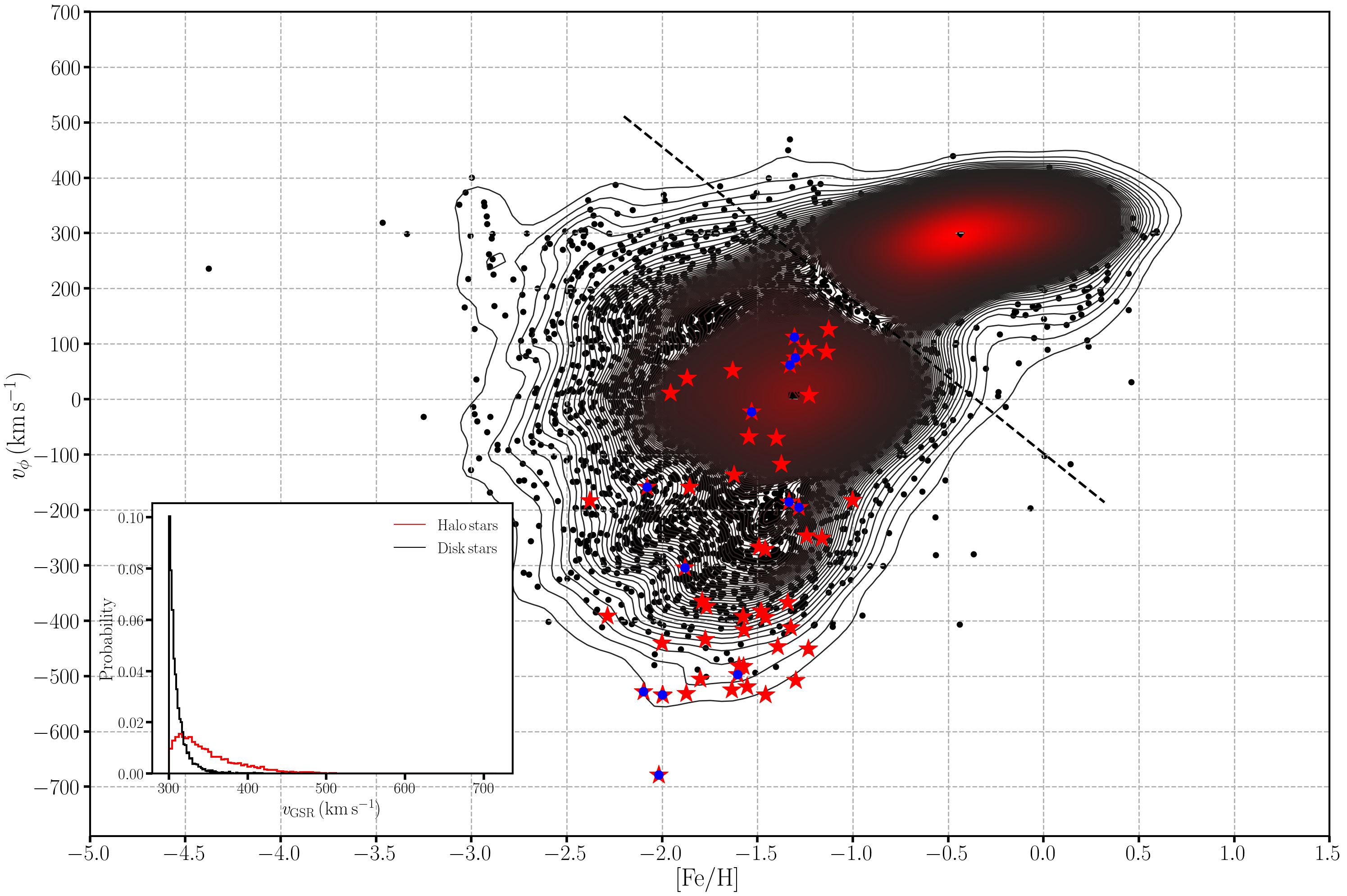}
\caption{The density distribution of $\it{v}_{\phi}$ vs. [Fe/H] for our HiVel candidates after the two selection steps. 
	The solid black lines represent the iso-density contours,
	and the black dotted line is the criterion we use to separate the disk population (upper right) and halo population (lower left). 
	The red stars mark the final HVS candidates discovered in this work, some of which (blue filled circles) were identified by \citet{Li2021}. 
	The inset shows the distributions of total velocity, $V_{\rm GSR}$, for the disk (black) and  halo (red) HiVel candidates.}
\end{center}
\end{figure*}

After nearly ten years of effort (2003--2013), 
the RAVE survey \citep{Steinmetz2006} collected 520,781 moderate-resolution ($R \sim 7500$) spectra centered on the Ca~I triplet (8410--8795\AA) for 457,588 unique stars, randomly selected from stars in the Southern Hemisphere with $9 < I < 12$,  using the multi-object spectrograph 6dF on the 1.2m UK Schmidt Telescope of the Australian Astronomical Observatory.
Estimates of line-of-sight velocities, stellar atmospheric parameters (effective temperature, $T_{\rm eff}$, surface gravity, $\log g$, and metallicity, [Fe/H]), as well as $\alpha$-element abundances for RAVE DR5 stars are described in \citet{Kunder2017}.
The typical uncertainties are 2\,km\,s$^{-1}$, 250\,K, 0.4\,dex, 0.2\,dex, and 0.2\,dex for $v_{\rm los}$, $T_{\rm eff}$, log\,$g$, [Fe/H], and [$\alpha$/Fe], respectively.

In this study, we also consider the data for over 800,000 stellar spectra (for over 700,000 stars) from SDSS\,DR12 \citep{Alam2015}, collected by the main SDSS survey,  SEGUE \citep{Yanny2009}, SEGUE-2 \citep{Rockosi2022}, and the Baryon Oscillation Spectroscopic Survey \citep[BOSS;][]{Dawson2013}. The $v_{\rm los}$, stellar atmospheric parameters, and [$\alpha$/Fe] abundance ratios are obtained from those spectra with the SEGUE Stellar Parameter Pipeline \citep[SSPP;][]{Lee2008a,Lee2011}.
Typical uncertainties are $5-10\,\rm{km\,s^{-1}}$,  $130\,\rm{K}$, $0.21\,\rm{dex}$, $0.11\,\rm{dex}$, and 0.10\,dex for $v_{\rm los}$, $T_{\rm eff}$, log\,$g$, [Fe/H], and [$\alpha$/Fe], respectively \citep{AllendePrieto2008,Lee2008b,Smolinski2011}.

LAMOST is a 4-meter quasi-meridian reflecting Schmidt telescope equipped with 4000 fibers distributed over a field of view $5^{\circ}$ in diameter \citep{Cui2012}. 
It can collect 4000 optical spectra per exposure, with wavelength coverage ranging from 3700 to 9000\,\AA\, and spectral resolving power around $R = 2000$. 
In the current work, we adopt the data from LAMOST\,DR8\footnote{\url{http://www.lamost.org/dr8/}}, which has released 6.5 million stellar spectra with reliable stellar parameter measurements for about 4 million unique stars.
The $v_{\rm los}$ and stellar atmospheric parameters and from LAMOST spectra are derived by the official pipeline -- LAMOST Stellar Parameter Pipeline \citep[LASP;][]{Luo2015}.
The typical uncertainties achieved by this pipeline are 5 \,km\,s$^{-1}$, 100\,K, 0.25\,dex, and 0.10\,dex for $v_{\rm los}$, $T_{\rm eff}$, log\,$g$, and [Fe/H], respectively. 
In addition to the parameters from the official pipeline, we also adopt the values of [$\alpha$/Fe] determined by \citet{Xiang2019}, with typical uncertainties of 0.05\,dex.

As an important part of SDSS-III/IV, the APOGEE survey \citep{Majewski2017} has obtained near-infrared ({\it H} band; 1.51-1.70\,$\mu$m) high-resolution ($R \sim$\,22500) spectra for 437,485 unique stars in the latest DR16 \citep{Ahumada2020}.
Estimates of  $v_{\rm los}$, the stellar atmospheric parameters, and 20 different elemental-abundance ratios are derived from APOGEE spectra by \citet{Jonsson2020}.
The typical uncertainties are about 0.5\,km\,s$^{-1}$, 100\,K, 0.10\,dex, 0.10\,dex, and 0.08\,dex for $v_{\rm los}$, $T_{\rm eff}$, log\,$g$, [Fe/H], and the individual elemental-abundance ratios, respectively.  

The GALAH survey is  a large-scale stellar spectroscopic survey aiming to collect optical (four discrete optical wavelength ranges: 4713--4903\,\AA\,, 5648--5873\,\AA\,, 6478--6737\,\AA\,, and 7585--7887\,\AA) high-resolution spectra ($R = $\,28,000) for around one million stars, using the HERMES spectrograph mounted on the 3.9\,m Anglo-Australian Telescope (AAT) at Siding Spring Observatory \citep{DeSilva2015}.
In the current work, we adopt information from GALAH DR2 \citep{Buder2018}, which contains estimates of $v_{\rm los}$, stellar atmospheric parameters, and 23 elemental abundances for 342,682 unique stars.  
The typical internal uncertainties are around 1.1\,km\,s$^{-1}$,
60\,K, 0.17\,dex, 0.10\,dex, and 0.02-0.10\,dex for $v_{\rm los}$, $T_{\rm eff}$, log\,$g$, [Fe/H],
and different elemental-abundance ratios, respectively, for FGK-type stars with a typical spectral signal-to-noise ratio (S/
N) around 40/1 per pixel.

The accuracy, especially the zero-point, of $v_{\rm los}$, is a key for selecting HiVel candidates in the current work.
We thus cross-match stars observed by the above large-scale surveys (except for SDSS) to the catalog of radial velocity standard stars constructed by \citet{Huang2018}.
The final adopted values of $v_{\rm los}$ are corrected for the zero-point offsets (listed in Table\,1) found using stars in common between those surveys and the catalog of radial velocity standard stars.
The targets of the SDSS survey are too faint to match with those of radial velocity standard stars; we adopt a $-7.3$\,km\,s$^{-1}$ offset for this correction from \citet{AdelmanMcCarthy2008}. 

\subsection{Astrometric Survey}
The European Space Agency (ESA) satellite {\it Gaia} \citep{GaiaCollaboration2016} recently released the Early Data Release 3 \citep[EDR3;][]{GaiaCollaboration2021}, which provides astrometric and photometric data for over 1.8 billion sources, with  $G$-band magnitude ranging from 3 to 21.
For parallax measurements, the typical uncertainties are 0.02-0.04\,mas, 0.07-0.1\,mas, and 0.5-1.4\,mas for $G < 15$, $= 17$, and $\sim 20-21$, respectively \citep{GaiaCollaboration2021}.
For proper-motion measurements,  the typical uncertainties are 0.02-0.04\,mas\,yr$^{-1}$, 0.07-0.1\,mas\,yr$^{-1}$, and 0.5-1.5\,mas\,yr$^{-1}$ for $G < 15$, $= 17$, and $\sim 20-21$, respectively \citep{GaiaCollaboration2021}.

\section{Selection of High-Velocity Stars}
\subsection{Coordinate Systems}
In this work, we adopt a right-handed Cartesian system, with $X$ toward the direction opposite to the Sun, $Y$ in the direction of Galactic rotation, and $Z$ in the direction of north Galactic pole (NGP), and a Galactocentric cylindrical system, with $R$ the projected Galactocentric distance, increasing radially outwards, $\phi$ toward the Galactic rotation direction, and $Z$ the same as that in the Cartesian system.
The three velocity components are represented by ($U$, $V$, $W$) in the Cartesian system and by ($V_{R}$, $V_{\phi}$, $V_{Z}$) in the Galactocentric cylindrical system, respectively.
We set the local standard of rest $(U_{\odot}, W_{\odot})=(7.01, 4.95)\,\rm{km\,s^{-1}}$ \citep{Huang2015}, and the $v_{\phi,\odot} = V_{\odot}+V_{c}(R_{0})$ is set to the value of $252.17$\,km\,s$^{-1}$, yielded by the proper motion of Sgr\,A$^{*}$ \citep{Reid2004} and the adopted value of Galactocentric distance of the Sun $R_0 = 8.34$\,kpc from \citep{Reid2014}.
The Sun is taken to be located above the disk, with $Z_{\odot}=25\,\rm{pc}$ \citep{BlandHawthorn2016}.
The main conclusions of this work still hold up if we adopt alternative values of $R_0$ (e.g., the most recent measurements by \citealt{GRAVITYCollaboration2019}), $Z_{\odot}$ \citep[e.g.,][]{Siegert2019} and solar motions \citep[e.g.,][]{Schonrich2012,Eilers2019,Zhou2022}.

\subsection{Distance and Total Velocity}
To derive the total velocities for our spectroscopic targets, accurate distances are required.
Rather than simply inverting the {\it Gaia} EDR3 parallax measurements, we estimate the distance from parallax by a Bayesian approach:
\begin{equation}
P(d|\varpi,\sigma_{\varpi})\propto P(\varpi|d,\sigma_{\varpi})d^{2}P(r)\text{,}
\end{equation}
where $r$ is the distance to the GC.

Similar to \citet{BailerJones2015}, the likelihood of parallax is given by:
\begin{equation}
P(\varpi|d,\sigma_{\varpi})=\frac{1}{\sqrt{2\pi}\sigma_{\varpi}}\exp\frac{-(\varpi-\varpi_{\rm zp}-\frac{1}{d})^{2}}{2\sigma^{2}_{\varpi}}\text{,}
\end{equation}
where $\varpi$ and $\sigma_{\varpi}$ are the parallax and its uncertainties from {\it Gaia} EDR3,
and $\varpi_{\rm zp}$ is the zero point of the parallax measurement, which is a function of ecliptic latitude, magnitude, and stellar color, and can be easily obtained by the procedure provided by \citet{Lindegren2021}. 

Similar to \citet{MCMillan2018}, the density prior $P (r)$ is defined as:
\begin{equation}
P(r)\propto N_{1}\exp(-\frac{R}{R_{1}}-\frac{|z|}{z_{1}})+N_{2}\exp(-\frac{R}{R_{2}}-\frac{|z|}{z_{2}})+N_{3}r^{-\alpha}\text{,}
\end{equation}
where $R_{1}$/$R_2$ and $z_{1}$/$z_2$ are the scale length and height of thin/thick disk, respectively.
The detailed values of those parameters are taken from \citet{MCMillan2018}.
Here $\alpha$ represents the power-law index of the inner-halo density profile, and is set to 3.39 \citep{Carollo2010}.
$N_1$, $N_2$, and $N_3$ are the normalization factors to ensure that the number density ratio of 0.15 \citep{Juric2008} between the thick and the thin disk, and of 0.005 \citep{Carollo2010} between the halo and the thin disk at the solar position. 

By using $v_{\rm los}$ from the aforementioned spectroscopic surveys, proper motions from {\it Gaia} EDR3, and the distances derived above, one can derive 3D velocities for all of our sample stars.
To do so, we cross-match the RAVE\,DR5, SDSS\,DR12, LAMOST\,DR8, APOGEE\,DR16, and GALAH\,DR2 targets to {\it Gaia} EDR3 using TOPCAT \citep{Taylor2005} with a  matching radius of 3 arcsec: 448,459, 712,742, 5,690,576, 428,876, and 339,890 sample stars are found, respectively.
The velocities in the Cartesian system (i.e., $U$, $V$, $W$) and the Galactocentric cylindrical system (i.e., $V_{R}$, $V_{\phi}$, $V_{Z}$) are calculated for all sample stars with reliable measurements of $v_{\rm los}$, proper motions, and distances. Uncertainties in these quantities are calculated by propagation of errors.
The total velocity $V_{\rm GSR}$ for each star can be easily derived by:
\begin{equation}
{V}_{\rm{GSR}} = (V_{R}^2 + V_{\phi}^2 + V_{z}^2)^{\frac{1}{2}}\text{.}
\end{equation}
In total, 84\% of the sample stars with reliable distance estimates ($\varpi \ge 0.2$\,mas and $\sigma_{\varpi}/\varpi \le 20$\%)  have their total velocities derived in the above manner.

\subsection{HiVel Star Selection}
With the total velocity calculated as above, we proceed to construct our HiVel sample. As a first step, we apply the following criteria to the above parent sample to select HiVel candidates with high-quality data:

\begin{enumerate}[label=\arabic*)]

\item $V_{\rm GSR} \ge 300$\,km\,s$^{-1}$;

\item Spectral S/N greater than 10 and uncertainties of $v_{\rm los}$ smaller than 30\,km\,s$^{-1}$;

\item Stellar atmospheric parameters ($T_{\rm eff}$, log\,$g$, and [Fe/H]) estimated from spectra;

\item $\varpi \ge 0.2$\,mas, $\sigma_{\varpi}$/$\varpi \leq$\,20\% and RUWE $< 1.4$;

\item Bad spectra are excluded, if found during our visual inspection.

\end{enumerate}

In total, 1406, 3449, 9786, 1264, and 1243 HiVel candidates are found from the RAVE, SDSS\,DR12, LAMOST, APOGEE, and GALAH surveys, respectively.
To exclude multiple entries, the star with the highest spectral S/N is kept if it was observed more than two times within a certain survey, and the entry with the highest spectral resolution (i.e., GALAH, APOGEE, RAVE, SDSS\,DR12, and finally LAMOST) is kept if the star was observed by two or more spectroscopic surveys. 
In this way, 14,894 unique stars are obtained.
For those 3651 stars with multiple observations, we check their $v_{\rm los}$ variations (the maximum minus the minimum); the result is shown in Fig.\,2.
A total of 756 stars with $v_{\rm los}$ variations larger than 15\,km\,s$^{-1}$ are marked as potential variable/binary stars, and are discarded from our HiVel sample.
Most recently, {\it Gaia}\,DR3 has released mean radial velocity measurements for 33 million stars with $G_{\rm RVS} < 14$\,mag \citep{Katz2022}. 
We thus check the radial velocity measurements of our HiVels from ground-based surveys to those from {\it Gaia} DR3.
In total, 5592 common stars are found.
The radial velocities measured from {\it Gaia} DR3 are consistent with those measured from the ground-based surveys, with a small offset of only $-0.45$\,km\,s$^{-1}$ (ground-based minus {\it Gaia}) and a tiny scatter of $3.26$\,km\,s$^{-1}$.
This check also clearly shows that the previous problems of radial-velocity determinations in {\it Gaia} DR2 due to unanticipated alignments between stars, as found by \citet{Boubert2019}, now have been largely resolved in {\it Gaia} DR3 \citep{Seabroke2021,Katz2022}.


\begin{figure}[!t]
\begin{center}
\includegraphics[scale=0.09,angle=0]{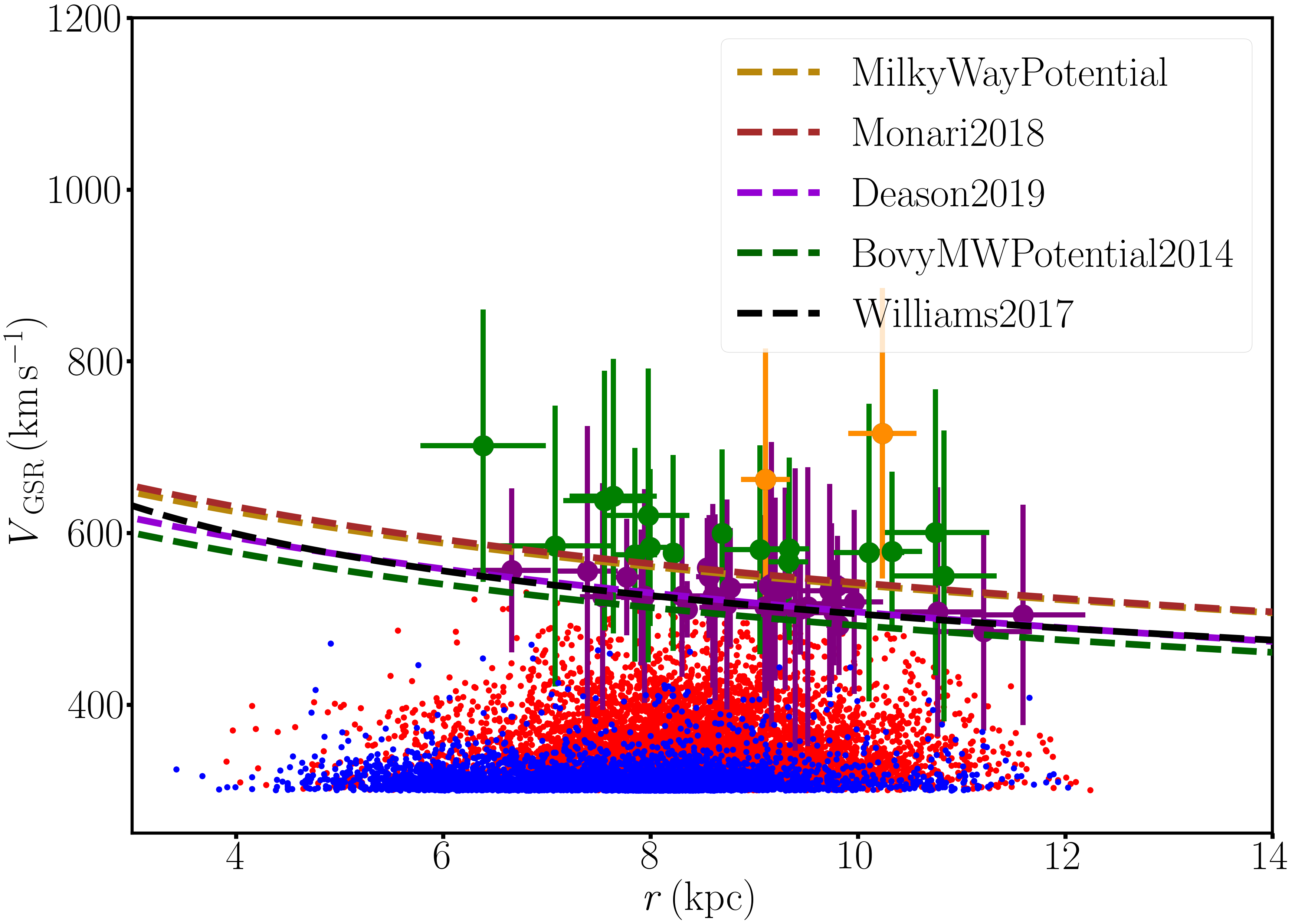}
\caption{Distribution of the sources of our final HiVel sample 
	in the plane of ${V}_{\rm{GSR}}$ and Galactocentric distance $r$. 
Red dots denote the halo population; blue dots, the disk population.
The dashed line is the Galactic escape velocity curve. 
The large dots with $\pm 1\sigma$ error bars represent the 52 HVS candidates with at least one group of $P_{\rm ub} \ge 50$\%, calculated from Monte Carlo simulations assuming the model Galactic escape velocity curves (dashed lines) derived from different Milky Way's potentials or the measured one (black dashed line), as indicated in the top-right corner.
The green and yellow large dots mark those candidates with all five groups of $P_{\rm ub}$ greater than 50 and 80 per cent, respectively.}
\end{center}
\end{figure}

\begin{figure*}[!t]
\begin{center}
\includegraphics[scale=0.22,angle=0]{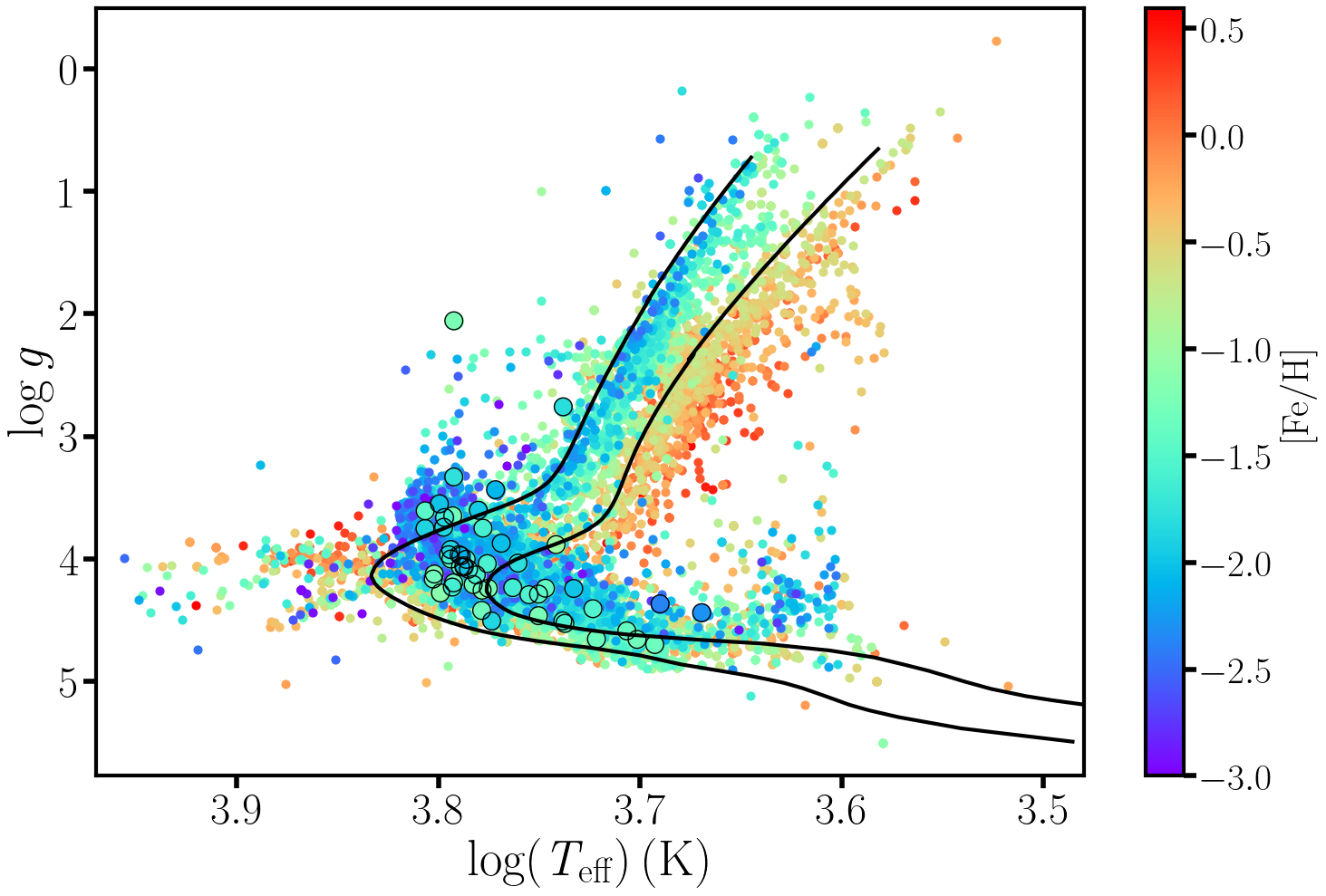}
\includegraphics[scale=0.22,angle=0]{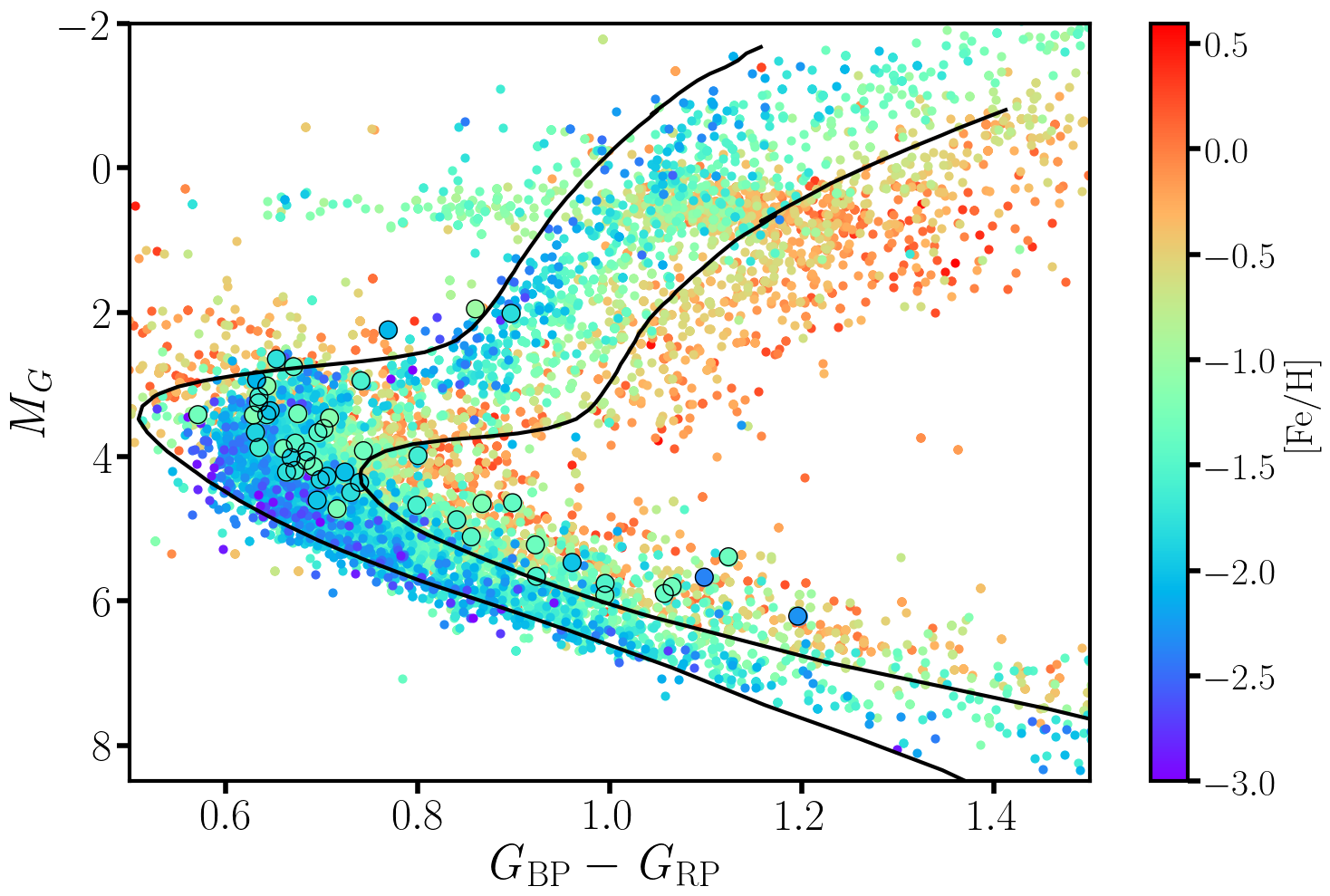}
\caption{The final HiVel sample shown in the log\,$g$--$T_{\rm eff}$ diagram (left panel) and 
	the diagram of $M_{\rm G_0}$  vs. color $(G_{\rm BP} - G_{\rm RP})_0$ (right panel). 
	The color is coded by metallicity, as labeled in the respective right sides. The larger dots are HVS candidates. The two solid lines in the panels represent stellar isochrones from PARSEC \citep{Bressan2012,Marigo2017}, 
with [M/H]\,$=-1.85$ and age\,$= 9$\,Gyr for the left line in each panel, and
[M/H]\,$=-0.60$ and age\,$= 14$\,Gyr for the right line in each panel.}
\end{center}
\end{figure*}

\begin{figure*}[!t]
\begin{center}
\includegraphics[scale=0.22,angle=0]{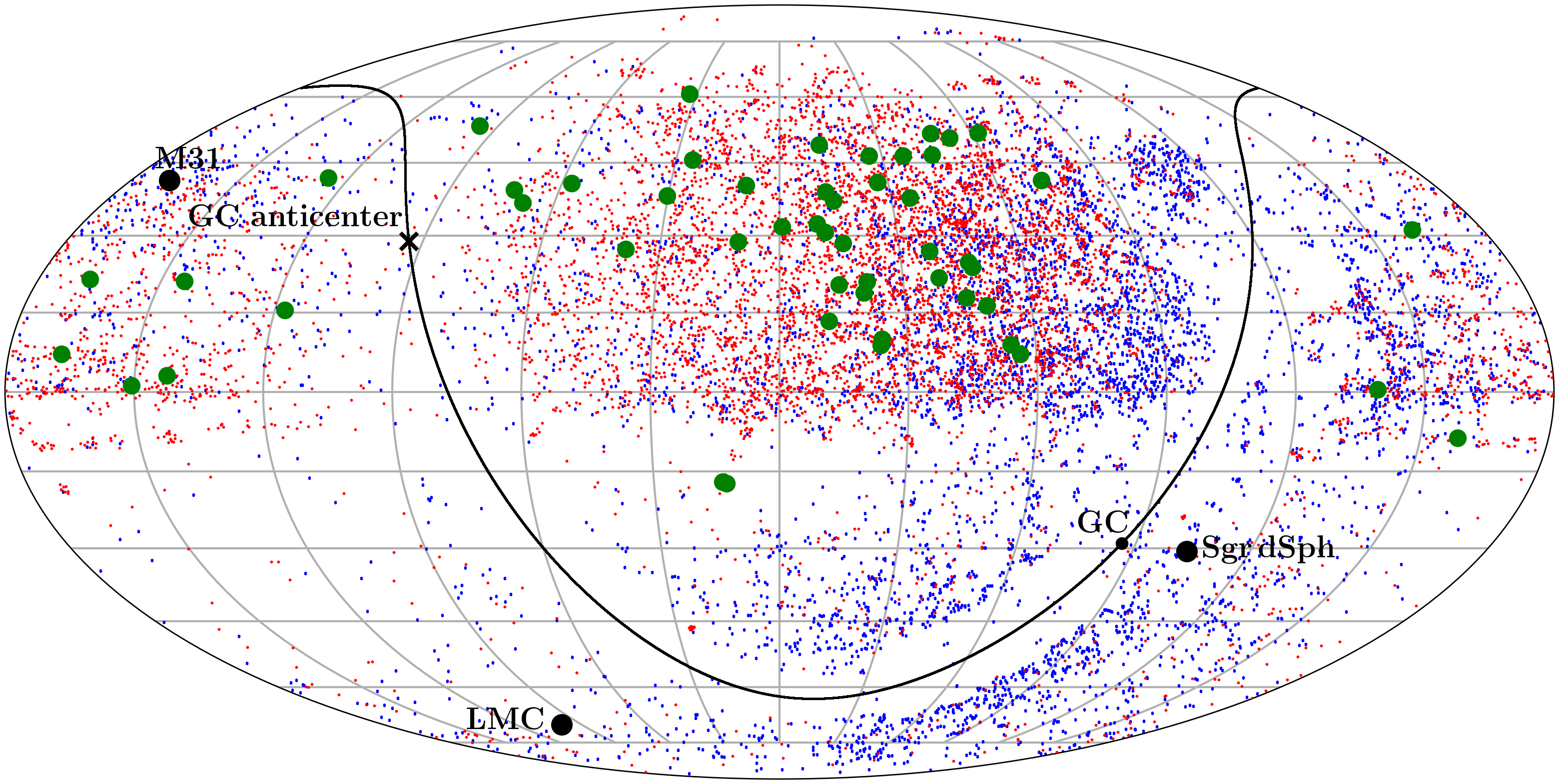}
\caption{Hammer projection in Right Ascension and Declination of the final HiVel stars. 
	Blue dots denote the disk population, red dots the halo population, and green solid circles indicate our HVS candidates.
The thick black line indicates the disk plane of the MW, 
with the large black dot and X-symbol marking the locations of the Galactic center and anti-center, respectively. 
The locations of M31, the LMC, and the Sgr\,dSph are also marked with their respective annotations.}
\end{center}
\end{figure*}

\begin{figure*}[!t]
\begin{center}
\includegraphics[scale=0.10,angle=0]{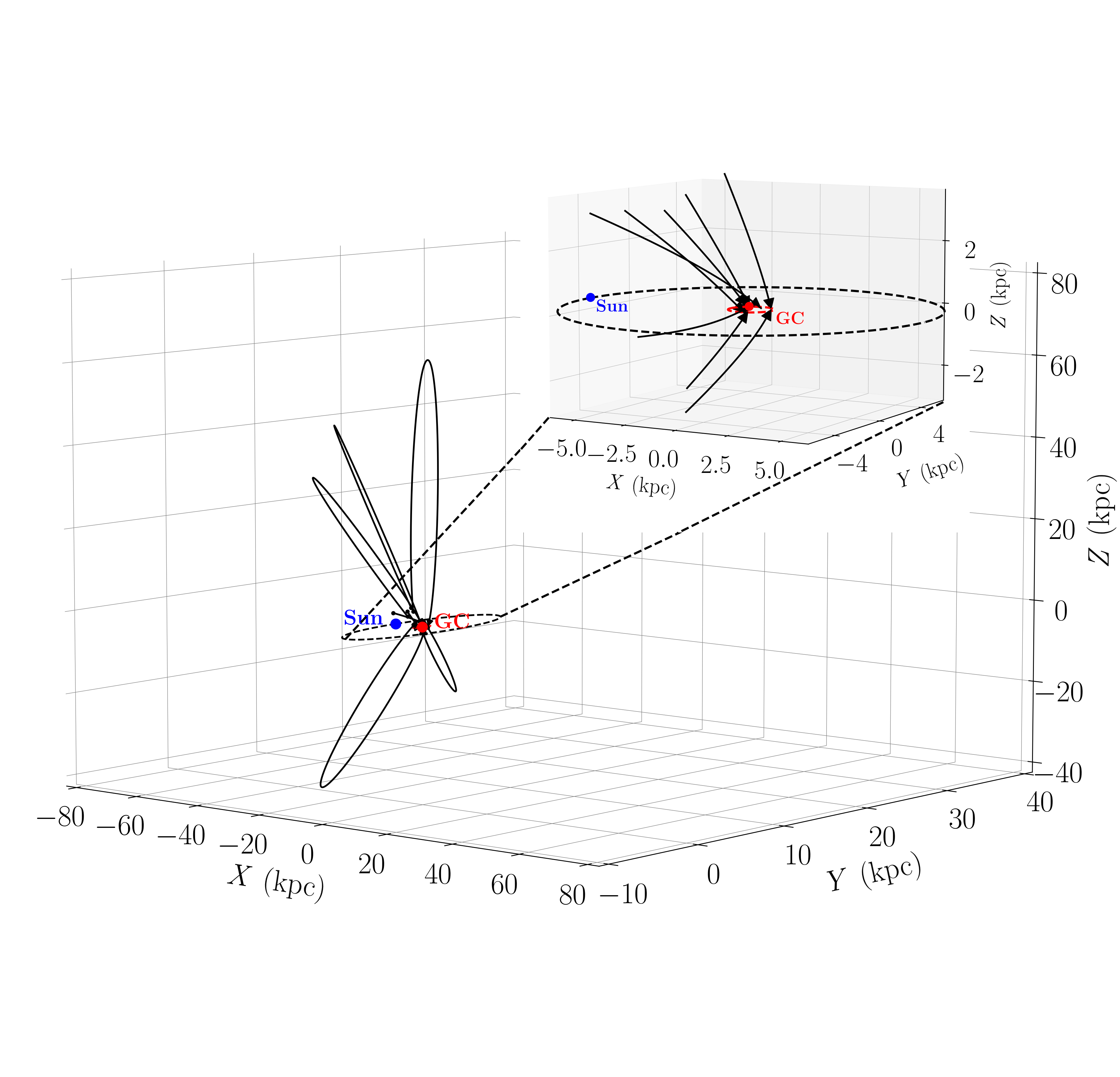}
\includegraphics[scale=0.08,angle=0]{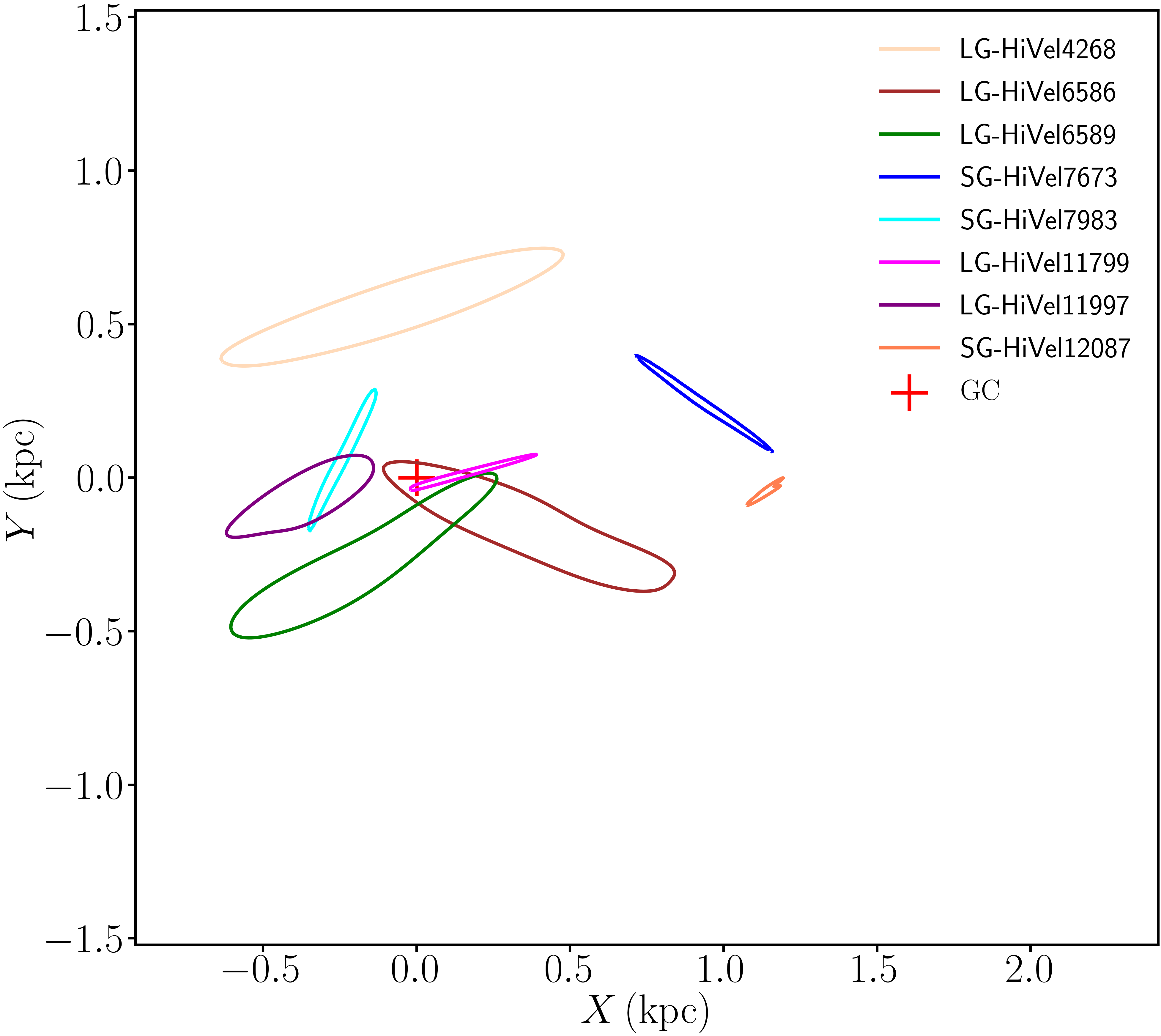}
\caption{{\it Left panel:} A 3D representation of the backward-integrated orbits of 8 HiVel stars passing within 1kpc near the GC.
The positions of the Sun and the GC are represented by the blue and red dots, respectively.
The Solar circle ($R$\,=\,8.34\,kpc) is represented by the dashed circle.
{\it Right panel:} 90\% confidence region of 8 HiVel stars in the Galactic plane ($X$--$Y$).}
\end{center}
\end{figure*}

\begin{figure*}[!t]
\begin{center}
\includegraphics[scale=0.2,angle=0]{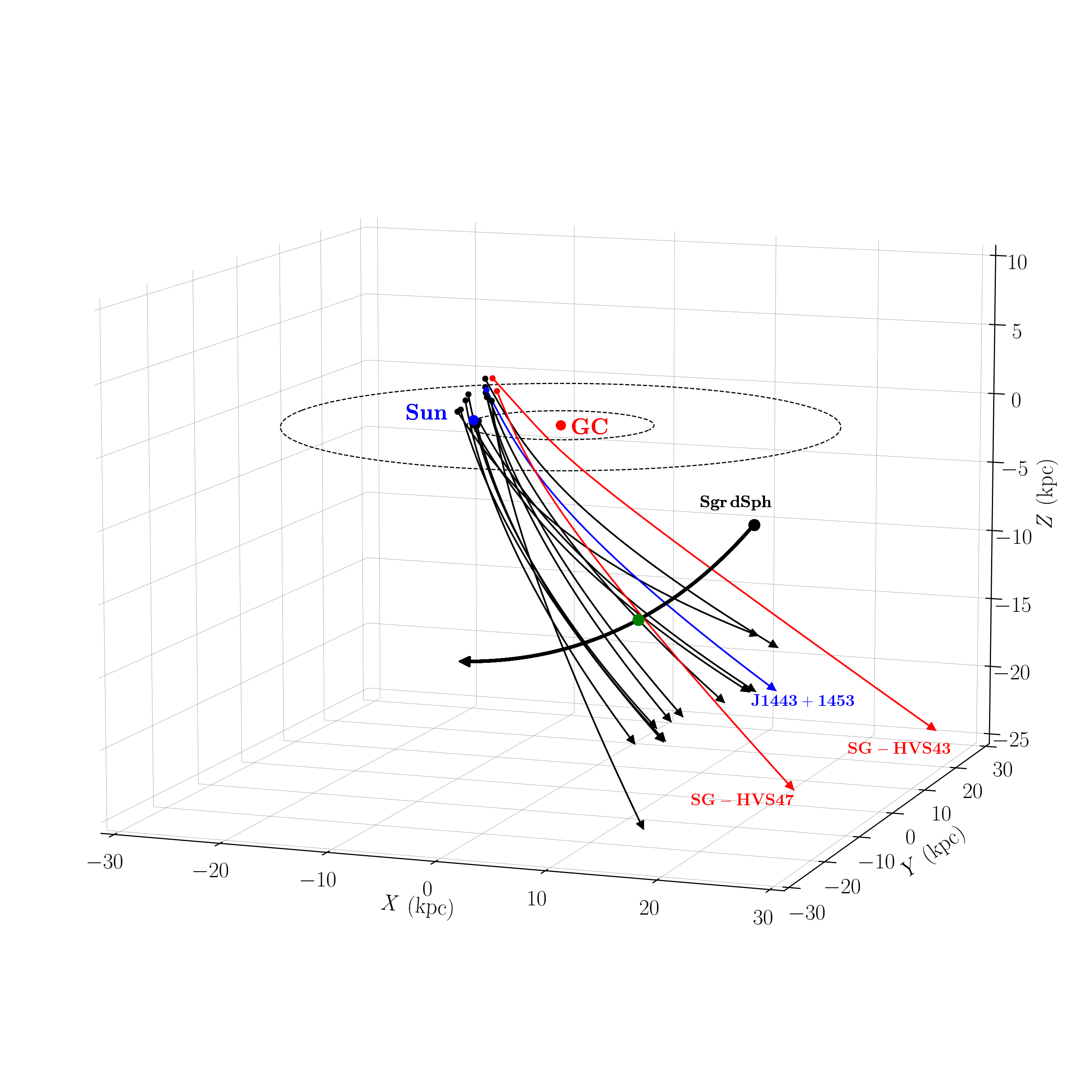}
\caption{A 3D representation of the backward-integrated orbits for the 15 extreme velocity stars and the Sgr dSph. 
The present positions of the stars, as well as the Sgr dSph, are marked as black dots. 
The green dot represents the latest pericenter of the Sgr dSph. The directions of the backward-integrated orbits are denoted by the arrows. 
The positions of the Sun and the Galactic Center are represented by blue and red dots, respectively. 
The Solar circle ($R$\,=\,8.34\,kpc) and the edge of the MW disk ($R$\,=\,25\,kpc) are denoted by the inner and outer dotted lines, respectively. 
The two red lines denote the backward-integrated orbits of two HVS candidates, SG-HVS43 and SG-HVS47; the blue line indicates the backward-integrated orbit of SG-HiVel6195 (also called J1443+1453), discovered by \citet{Huang2021}, the first HVS candidate originating from the Sgr dSph.}
\end{center}
\end{figure*}

\begin{figure*}[!t]
\begin{center}
\includegraphics[scale=0.38,angle=0]{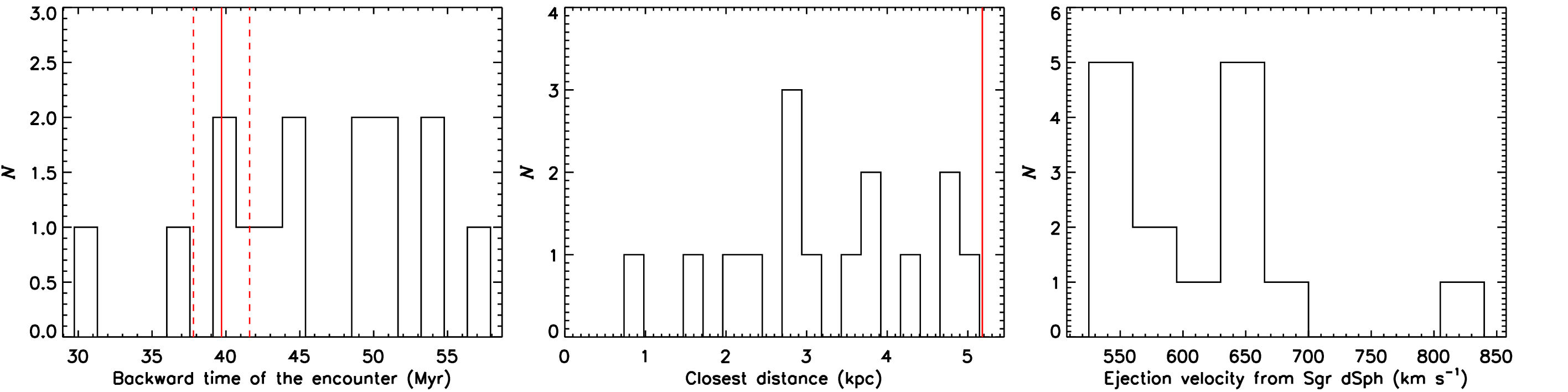}
\caption{{\it Left panel:} The distribution of backward time for the 15 extreme velocity stars that have close encounters with the Sgr dSph within its $2 r_h$. 
	The red solid line indicates the backward time of the Sgr dSph from the present position to its latest pericenter (39.7 Myr), and the red dashed lines mark its $1\sigma$ uncertainty (calculated using the Monte Carlo simulation similar to that described in Section 5.2). 
{\it Middle panel:} The distribution of the closest distances between the 15 stars and the Sgr dSph. 
The red solid line denotes two times $r_h$ of the Sgr dSph.
{\it Right panel:} The distribution of the ejection velocities of the 15 stars, assuming their Sgr dSph origin.}
\end{center}
\end{figure*}

Second, we search for contamination from, e.g., variable stars and white dwarfs with poor $v_{\rm los}$ determinations.
We cross-match the HiVel sample to the existing catalogs of variable stars, including the General Catalogue of Variable Stars \citep{Samus2017}, the Catalina variable catalogs \citep{Drake2013b,Drake2013a,Drake2014,Drake2017,Torrealba2015},  the ASAS-SN catalogs of variable stars \citep{Shappee2014,Jayasinghe2018,Jayasinghe2019a,Jayasinghe2019b}, and the {\it Gaia} DR2 catalog of variable stars \citep{GaiaCollaboration2019}; 289 potential variable stars are excluded from the HiVel sample.
In addition, 208 potential white dwarf candidates \citep[by matching with the white dwarf catalog from {\it Gaia} DR2;][]{GentileFusillo2019} in our sample are removed, since almost all of them have very poor $v_{\rm los}$ determinations. Two potential M31/M33 member stars are also removed from our HiVel sample.


Following the above two steps, 
over 13,000 HiVel candidates with total velocities $V_{\rm GSR} \ge 300$\,km\,s$^{-1}$ remain. 
We plot them on the $V_{\phi}$--[Fe/H] diagram (Fig.\,3) to help understand their parent populations.
From inspection, there are clearly at least two main clumps:
the halo population clustered at low metallicity and $V_{\phi} \sim 0$\,km\,s$^{-1}$, 
and the disk population clustered at high metallicity and $V_{\phi} = 200$--$300$\,km\,s$^{-1}$.
We apply an empirical demarcation line, $V_{\phi} =- 276.70\times{\rm [Fe/H]} - 97.78$, 
and classify the candidates into the two populations (disk and halo).
With this classification, we then re-calculate the distance estimate for each star 
by updating the density prior (keep the first two terms for a disk star, 
and only the last term for a halo star),
 and then re-estimate the total velocity.
Then we again apply the HiVel criterion ($V_{\rm GSR} \ge 300$\,km\,s$^{-1}$) to the candidates.
By iterating the above processes twice, the distance estimate, total velocity calculation, and disk/halo classification converge.

Finally, 12,784 HiVel candidates are selected as our HiVel sample, with 6966 classified as halo stars and 5818 classified as disk stars. As shown in Fig.\,3, the total velocity distributions of halo and disk HiVel stars are quite different.  
While the number of the disk population drops quickly to nearly zero at $V_{\rm GSR} = 350$\,km\,s$^{-1}$, 
the halo population exhibits a long-tailed distribution, extending beyond $-$500 to $-$700\,km\,s$^{-1}$ in $v_{\phi}$. 

To select potential HVS, four models of the Galactic potential, two from \citet{Monari2018} and \citet{Deason2019}, and two from the {\tt MilkyWayPotential} and {\tt BovyMWPotential2014} potentials implemented in {\tt Gala} \citep{PriceWhelan2017},  are adopted to estimate the Galactic escape velocities.
In addition, one measured escape velocity curve from \citet{Williams2017} is adopted.
For each of the HiVels, we perform 2000 Monte Carlo simulations for estimating $V_{\rm{GSR}}$ by sampling the measurement errors (assuming Gaussian distributions, except for the distance whose distributions directly given by Equation\,1).
The unbound probability  $P_{\rm ub}$ of each star is then calculated by counting the frequency of its $V_{\rm GSR}$ exceeding the Galactic escape velocity curves in the 2000 Monte Carlo simulations.
In this way, five groups of $P_{\rm ub}$ can be obtained for each star using the five Galactic escape velocity curves mentioned above. As shown in Fig.\,4 and Table 2, a total of 52 HVS candidates with at least one group of $P_{\rm ub} \ge 50$\% were found \citep[40 for the first time, and 12 reported by][]{Bromley2018,Li2021}. We note that these 52 stars are only marginally unbound to the Milky Way's potential. 
Most of them have unbound probabilities smaller than $1\sigma$ and only 6 stars show $\it{P}_{\rm{ub}} > 68$\% if adopting the Milky Way potential with the largest escape velocities, i.e. {\tt MilkyWayPotential}. 
Among these, 19 candidates have all five groups of $P_{\rm ub} \ge 50$\% \citep[5 reported by][]{Li2021}.
More interestingly, all of these candidates are metal-poor $\alpha$-enhanced halo stars 
(see the insets of Fig.\,1 and Fig.\,3). Both the log\,$g$--$T_{\rm eff}$ diagram 
and the diagram of absolute magnitude ($M_G$) vs. color ($G_{\rm BP} - G_{\rm RP}$)   
further indicate that most of those HVS candidates are old turn-off stars with ages roughly between 9--14\,Gyr (see Fig.\,5).  

Detailed information for our newly discovered HVS candidates, as well as those collected from the literature (numbering 88 in total), is listed in Table\,A1.

\begin{table*}[!b]
\centering
\begin{threeparttable}
\tablename{ 2: The 52 HVS candidates assuming different escape velocity curves}
\label{tbl:table2}
\resizebox{\textwidth}{!}{
\begin{tabular}{llllllllllll}
\toprule
\multirow{1}{*} Notation & $\rm{ra}$ & $\mu_{\alpha}\rm{cos}\,\delta$ & $\mu_{\delta}$ & $v_{\rm{los}}$ & $d$ & $V_{\rm{GSR}}$  & $P_{\rm ub}$ (MW)\tnote{a} & $P_{\rm ub}$ (M18)\tnote{b}&$P_{\rm ub}$ (D19)\tnote{c} & $P_{\rm ub}$ (BMW)\tnote{d} & $P_{\rm ub}$ (W17)\tnote{e} \\ 
Gaia ID & $\rm{dec}$ & $\sigma_{\mu_{\alpha}\rm{cos}\,\delta}$ & $\sigma_{\mu_{\delta}}$  & $\sigma_{v_{\rm{los}}}$ & $\sigma_{d}$ & $\sigma_{V_{\rm{GSR}}}$ & & & & & \\
&  $\rm{(deg)}$ & $\rm{(mas\,yr^{-1})}$ & $\rm{(mas\,yr^{-1})}$ & $\rm{(km\,s^{-1})}$ & $\rm{(kpc)}$ & $\rm{(km\,s^{-1})}$ &   &   &   &   &   \\
\cline{1-12}
\multirow{1}{*} LG-HVS1
& 12.362361 & 11.129 & $-$59.488 & $-$345.370 & 2.328 & 535.012 & 0.46 & 0.45 & 0.58 & 0.65 & 0.59 \\
2556679991337823488 & 7.128955 & 0.087 & 0.078 & 9.480 & 0.467 & 117.951 & & & & & \\
\multirow{1}{*} LG-HVS2
& 12.509631 & 48.314 & 1.221 & $-$32.940 & 2.711 & 517.589 & 0.40 & 0.39 & 0.54 & 0.61 & 0.55 \\
2801887851883799936 & 21.420701 & 0.065 & 0.039 & 14.380 & 0.438 & 93.602 & & & & & \\
\multirow{1}{*} SG-HVS3
& 29.303224 & $-$15.182 & $-$38.759 & 217.004 & 2.458 & 507.086 & 0.30 & 0.29 & 0.49 & 0.59 & 0.50 \\
2510946771548268160 & 1.193651 & 0.071 & 0.056 & 2.256 & 0.426 & 61.543 & & & & & \\
\multirow{1}{*} LG-HVS4
& 35.653263 & 75.919 & 10.855 & 151.050 & 1.515 & 511.539 & 0.27 & 0.26 & 0.48 & 0.61 & 0.49 \\
87667257538859264 & 21.079511 & 0.089 & 0.067 & 9.470 & 0.183 & 53.377 & & & & & \\
\multirow{1}{*} LG-HVS5
& 37.488910 & 57.169 & $-$40.756 & $-$263.450 & 2.764 & 716.088 & 0.92 & 0.91 & 0.96 & 0.97 & 0.96 \\
2503400067332349440 & 3.130835 & 0.065 & 0.055 & 18.680 & 0.539 & 169.393 & & & & & \\
\multirow{1}{*} LG-HVS6
& 54.010210 & 32.181 & $-$26.060 & 67.550 & 3.576 & 504.377 & 0.45 & 0.44 & 0.55 & 0.60 & 0.55 \\
237369721329578112 & 41.872442 & 0.053 & 0.039 & 13.920 & 0.697 & 128.468 & & & & & \\
\multirow{1}{*} SG-HVS7
& 62.335793 & $-$21.928 & $-$86.078 & 16.314 & 1.519 & 532.706 & 0.48 & 0.47 & 0.58 & 0.64 & 0.59 \\
45249542048091392 & 15.531954 & 0.182 & 0.125 & 3.303 & 0.320 & 124.369 & & & & & \\
\multirow{1}{*} LG-HVS8
& 84.087702 & 75.773 & $-$54.534 & $-$118.040 & 1.596 & 492.411 & 0.22 & 0.21 & 0.41 & 0.51 & 0.42 \\
263868123355334784 & 53.192993 & 0.049 & 0.040 & 14.120 & 0.133 & 57.819 & & & & & \\
\multirow{1}{*} SG-HVS9
& 107.714100 & $-$39.266 & $-$70.707 & $-$43.470 & 1.711 & 519.621 & 0.43 & 0.42 & 0.54 & 0.59 & 0.54 \\
946735552249457152 & 39.286968 & 0.091 & 0.080 & 1.811 & 0.300 & 107.155 & & & & & \\
\multirow{1}{*} SG-HVS10
& 111.533120 & 46.016 & $-$46.600 & $-$118.499 & 2.556 & 600.593 & 0.68 & 0.67 & 0.77 & 0.81 & 0.77 \\
897267428899789440 & 36.689452 & 0.080 & 0.064 & 1.912 & 0.563 & 166.708 & & & & & \\
\multirow{1}{*} SG-HVS11
& 122.591610 & $-$0.332 & $-$59.917 & $-$33.970 & 2.804 & 549.726 & 0.54 & 0.54 & 0.63 & 0.67 & 0.64 \\
921561993012731648 & 40.652329 & 0.065 & 0.043 & 3.204 & 0.599 & 169.475 & & & & & \\
\multirow{1}{*} SG-HVS12
& 141.546960 & $-$17.158 & $-$33.579 & 223.547 & 3.767 & 484.930 & 0.37 & 0.36 & 0.47 & 0.52 & 0.47 \\
694458695225567872 & 27.340210 & 0.051 & 0.037 & 2.035 & 0.699 & 115.507 & & & & & \\
\multirow{1}{*} SG-HVS13
& 147.216280 & 11.123 & $-$128.292 & 75.051 & 1.177 & 514.638 & 0.22 & 0.20 & 0.47 & 0.63 & 0.50 \\
1050729011271507328 & 60.705308 & 0.039 & 0.051 & 2.637 & 0.069 & 39.607 & & & & & \\
\multirow{1}{*} SG-HVS14
& 149.660010 & 39.411 & $-$23.042 & 38.188 & 2.849 & 578.540 & 0.69 & 0.67 & 0.83 & 0.88 & 0.84 \\
803054228887217024 & 38.134845 & 0.050 & 0.045 & 3.035 & 0.469 & 92.671 & & & & & \\
\multirow{1}{*} LG-HVS15
& 154.757012 & $-$17.830 & $-$40.305 & $-$134.730 & 3.542 & 507.821 & 0.43 & 0.42 & 0.53 & 0.57 & 0.53 \\
809462835487742080 & 45.733809 & 0.044 & 0.055 & 19.400 & 0.715 & 145.156 & & & & & \\
\multirow{1}{*} SG-HVS16
& 166.340900 & 43.549 & $-$18.727 & 313.306 & 2.397 & 599.505 & 0.71 & 0.69 & 0.84 & 0.90 & 0.85 \\
3559325645434651648 & $-$17.070308 & 0.088 & 0.064 & 2.566 & 0.461 & 97.656 & & & & & \\
\multirow{1}{*} SG-HVS17
& 167.381140 & $-$24.318 & $-$36.315 & 439.254 & 3.754 & 662.136 & 0.82 & 0.81 & 0.90 & 0.93 & 0.91 \\
3559089525313289344 & $-$17.284739 & 0.055 & 0.045 & 3.208 & 0.795 & 152.825 & & & & & \\
\multirow{1}{*} LG-HVS18
& 169.502405 & $-$36.101 & 4.092 & 157.240 & 3.097 & 537.966 & 0.50 & 0.48 & 0.71 & 0.81 & 0.73 \\
3998883554967849216 & 28.818153 & 0.050 & 0.081 & 16.140 & 0.392 & 58.385 & & & & & \\
\multirow{1}{*} LG-HVS19
& 170.792361 & $-$35.643 & $-$35.437 & 34.180 & 3.359 & 577.195 & 0.59 & 0.59 & 0.68 & 0.72 & 0.69 \\
770479307125812864 & 40.248147 & 0.055 & 0.058 & 15.530 & 0.741 & 173.473 & & & & & \\
\multirow{1}{*} LG-HVS20
& 180.581777 & $-$36.995 & $-$32.226 & 26.330 & 3.191 & 513.930 & 0.41 & 0.40 & 0.50 & 0.54 & 0.50 \\
4026489543162246912 & 31.837635 & 0.074 & 0.055 & 19.690 & 0.711 & 162.690 & & & & & \\
\multirow{1}{*} SG-HVS21
& 189.653260 & $-$43.531 & $-$7.011 & $-$66.739 & 3.196 & 534.394 & 0.45 & 0.43 & 0.56 & 0.63 & 0.57 \\
1514459756956334720 & 32.344768 & 0.057 & 0.056 & 3.493 & 0.547 & 106.635 & & & & & \\
\multirow{1}{*} LG-HVS22
& 191.641073 & $-$23.127 & $-$33.598 & $-$88.910 & 3.917 & 510.819 & 0.42 & 0.41 & 0.49 & 0.53 & 0.50 \\
1513259266353444992 & 30.695884 & 0.049 & 0.062 & 11.830 & 0.860 & 164.598 & & & & & \\
\multirow{1}{*} SG-HVS23
& 191.669490 & 27.761 & $-$30.708 & 360.732 & 1.936 & 510.891 & 0.13 & 0.12 & 0.34 & 0.53 & 0.37 \\
3929069346904353152 & 13.430034 & 0.081 & 0.075 & 1.581 & 0.264 & 32.834 & & & & & \\
\multirow{1}{*} SG-HVS24
& 191.991830 & $-$41.337 & $-$51.052 & 81.992 & 2.421 & 540.108 & 0.49 & 0.48 & 0.56 & 0.60 & 0.57 \\
1567338780126678400 & 48.967882 & 0.062 & 0.080 & 1.482 & 0.558 & 165.635 & & & & & \\
\multirow{1}{*} SG-HVS25
& 192.498020 & $-$52.733 & $-$67.243 & $-$168.672 & 1.900 & 536.101 & 0.41 & 0.39 & 0.60 & 0.69 & 0.62 \\
1520968079814522496 & 38.947287 & 0.030 & 0.032 & 2.317 & 0.176 & 69.351 & & & & & \\
\multirow{1}{*} LG-HVS26
& 194.347116 & $-$7.334 & $-$51.745 & 175.760 & 2.876 & 526.631 & 0.39 & 0.38 & 0.50 & 0.57 & 0.51 \\
3942455900972145152 & 20.419876 & 0.069 & 0.052 & 12.660 & 0.466 & 107.150 & & & & & \\
\multirow{1}{*} LG-HVS27
& 194.385615 & $-$16.537 & $-$43.399 & $-$221.790 & 3.587 & 581.297 & 0.64 & 0.62 & 0.77 & 0.82 & 0.78 \\
1517369756214955008 & 37.110194 & 0.023 & 0.026 & 15.930 & 0.510 & 106.325 & & & & & \\
\multirow{1}{*} LG-HVS28
& 196.007951 & $-$5.770 & $-$38.240 & $-$154.570 & 3.847 & 514.009 & 0.37 & 0.36 & 0.49 & 0.56 & 0.50 \\
1461216666591966336 & 28.552378 & 0.037 & 0.051 & 11.840 & 0.620 & 106.636 & & & & & \\
\multirow{1}{*} LG-HVS29
& 200.292967 & $-$74.467 & $-$56.547 & $-$58.030 & 1.833 & 576.624 & 0.56 & 0.54 & 0.68 & 0.75 & 0.69 \\
3938867537399911680 & 18.835966 & 0.088 & 0.051 & 10.990 & 0.261 & 114.295 & & & & & \\
\multirow{1}{*} LG-HVS30
& 201.229653 & $-$31.015 & $-$40.112 & 380.200 & 2.554 & 526.695 & 0.38 & 0.37 & 0.50 & 0.57 & 0.51 \\
1442286920356579712 & 20.939786 & 0.085 & 0.057 & 7.650 & 0.572 & 94.598 & & & & & \\
\multirow{1}{*} LG-HVS31
& 203.794513 & $-$43.676 & $-$9.073 & 238.430 & 3.071 & 524.692 & 0.32 & 0.31 & 0.46 & 0.55 & 0.48 \\
3725682648069934336 & 8.899663 & 0.049 & 0.028 & 14.730 & 0.430 & 79.089 & & & & & \\
\multirow{1}{*} SG-HVS32
& 204.084080 & $-$45.700 & $-$6.408 & 33.257 & 3.563 & 620.146 & 0.65 & 0.64 & 0.73 & 0.77 & 0.73 \\
3726042154012422656 & 10.005512 & 0.068 & 0.048 & 2.925 & 0.815 & 171.275 & & & & & \\
\multirow{1}{*} LG-HVS33
& 206.252711 & 15.691 & $-$29.640 & $-$104.200 & 3.748 & 566.398 & 0.61 & 0.60 & 0.78 & 0.86 & 0.79 \\
1503968702337282048 & 46.596244 & 0.030 & 0.038 & 16.340 & 0.539 & 78.929 & & & & & \\
\multirow{1}{*} SG-HVS34
& 207.085230 & 16.371 & $-$54.602 & 216.560 & 2.027 & 559.244 & 0.52 & 0.50 & 0.75 & 0.86 & 0.78 \\
1500305198313028224 & 40.949577 & 0.048 & 0.060 & 1.558 & 0.283 & 58.227 & & & & & \\
\multirow{1}{*} LG-HVS35
& 214.976283 & $-$42.857 & $-$17.186 & $-$245.320 & 3.358 & 549.047 & 0.47 & 0.45 & 0.64 & 0.74 & 0.66 \\
1484524973071001984 & 37.669366 & 0.023 & 0.029 & 14.680 & 0.345 & 71.489 & & & & & \\
\multirow{1}{*} SG-HVS36
& 216.185130 & $-$27.628 & $-$20.628 & $-$338.217 & 4.151 & 506.896 & 0.26 & 0.25 & 0.45 & 0.57 & 0.47 \\
1506678826700636544 & 46.550514 & 0.020 & 0.024 & 1.402 & 0.395 & 58.074 & & & & & \\
\multirow{1}{*} SG-HVS37
& 217.396790 & $-$3.753 & $-$54.087 & $-$112.495 & 2.912 & 582.995 & 0.61 & 0.59 & 0.74 & 0.81 & 0.75 \\
1280443412952917632 & 26.915099 & 0.040 & 0.039 & 2.455 & 0.371 & 91.517 & & & & & \\
\cline{1-12}
\end{tabular}}
\end{threeparttable}
\end{table*}

\begin{table*}[!t]
\centering
\begin{threeparttable}
\tablename{ 2: The 52 HVS candidates assuming different escape velocity curves}
\label{tbl:table2}
\resizebox{\textwidth}{!}{
\begin{tabular}{llllllllllll}
\toprule
\multirow{1}{*} Notation & $\rm{ra}$ & $\mu_{\alpha}\rm{cos}\,\delta$ & $\mu_{\delta}$ & $v_{\rm{los}}$ & $d$ & $V_{\rm{GSR}}$  & $P_{\rm ub}$ (MW)\tnote{a} & $P_{\rm ub}$ (M18)\tnote{b}& $P_{\rm ub}$ (D19)\tnote{c}& $P_{\rm ub}$ (MMW)\tnote{d} & $P_{\rm ub}$ (W17)\tnote{e} \\ 
Gaia ID & $\rm{dec}$ & $\sigma_{\mu_{\alpha}\rm{cos}\,\delta}$ & $\sigma_{\mu_{\delta}}$  & $\sigma_{v_{\rm{los}}}$ & $\sigma_{d}$ & $\sigma_{V_{\rm{GSR}}}$ & & & & & \\
&  $\rm{(deg)}$ & $\rm{(mas\,yr^{-1})}$ & $\rm{(mas\,yr^{-1})}$ & $\rm{(km\,s^{-1})}$ & $\rm{(kpc)}$ & $\rm{(km\,s^{-1})}$ &   &   &   &   &   \\
\cline{1-12}
\multirow{1}{*} SG-HVS38
& 218.817840 & $-$47.200 & $-$36.613 & $-$237.299 & 2.679 & 548.674 & 0.41 & 0.39 & 0.61 & 0.71 & 0.62 \\
1242022529603565440 & 21.728278 & 0.026 & 0.031 & 1.963 & 0.256 & 67.944 & & & & & \\
\multirow{1}{*} LG-HVS39
& 224.735948 & 7.222 & $-$29.369 & $-$186.540 & 4.389 & 580.327 & 0.60 & 0.59 & 0.71 & 0.78 & 0.72 \\
1587301410160177920 & 46.728570 & 0.046 & 0.054 & 13.830 & 0.905 & 121.657 & & & & & \\
\multirow{1}{*} LG-HVS40
& 224.807592 & $-$16.170 & $-$65.173 & $-$66.600 & 2.279 & 525.834 & 0.39 & 0.38 & 0.48 & 0.54 & 0.49 \\
1188512524199896064 & 17.866868 & 0.063 & 0.087 & 15.610 & 0.426 & 131.915 & & & & & \\
\multirow{1}{*} LG-HVS41
& 226.584048 & $-$2.004 & $-$43.521 & $-$131.440 & 3.855 & 642.758 & 0.72 & 0.72 & 0.80 & 0.84 & 0.81 \\
1264622956753148416 & 24.765407 & 0.036 & 0.050 & 16.960 & 0.803 & 160.073 & & & & & \\
\multirow{1}{*} SG-HVS42
& 227.154450 & 3.842 & $-$30.310 & $-$211.830 & 4.091 & 538.205 & 0.46 & 0.45 & 0.57 & 0.63 & 0.58 \\
1592830751057210496 & 51.533092 & 0.051 & 0.058 & 2.178 & 0.906 & 122.536 & & & & & \\
\multirow{1}{*} SG-HVS43
& 227.350790 & $-$42.685 & 11.386 & 38.323 & 3.395 & 637.638 & 0.71 & 0.70 & 0.80 & 0.84 & 0.80 \\
1263758598878844288 & 23.835252 & 0.059 & 0.072 & 2.909 & 0.780 & 151.478 & & & & & \\
\multirow{1}{*} SG-HVS44
& 229.492430 & $-$38.172 & $-$15.989 & $-$269.744 & 3.887 & 584.788 & 0.53 & 0.53 & 0.61 & 0.67 & 0.62 \\
1208146095315810944 & 16.324317 & 0.069 & 0.056 & 4.129 & 0.905 & 163.723 & & & & & \\
\multirow{1}{*} SG-HVS45
& 232.301250 & $-$5.243 & $-$42.533 & $-$189.458 & 3.008 & 513.102 & 0.36 & 0.36 & 0.47 & 0.54 & 0.49 \\
1594746199095755520 & 50.496393 & 0.063 & 0.075 & 3.028 & 0.560 & 108.323 & & & & & \\
\multirow{1}{*} LG-HVS46
& 234.156763 & 5.004 & $-$38.522 & 16.630 & 4.504 & 701.442 & 0.80 & 0.79 & 0.88 & 0.91 & 0.89 \\
1164837294370596992 & 9.002340 & 0.043 & 0.044 & 8.130 & 0.900 & 158.662 & & & & & \\
\multirow{1}{*} SG-HVS47
& 236.220050 & $-$39.882 & 13.278 & 87.606 & 2.953 & 556.258 & 0.42 & 0.40 & 0.54 & 0.61 & 0.55 \\
4429852530637477760 & 7.130787 & 0.064 & 0.051 & 4.646 & 0.545 & 95.418 & & & & & \\
\multirow{1}{*} SG-HVS48
& 241.981170 & $-$14.429 & $-$32.566 & $-$282.395 & 3.724 & 516.436 & 0.39 & 0.38 & 0.49 & 0.54 & 0.49 \\
1403645069530607616 & 51.582219 & 0.063 & 0.082 & 3.099 & 0.760 & 122.287 & & & & & \\
\multirow{1}{*} SG-HVS49
& 252.688090 & $-$6.120 & $-$38.142 & $-$126.759 & 3.373 & 526.380 & 0.40 & 0.39 & 0.49 & 0.56 & 0.50 \\
1356298174692911104 & 41.324670 & 0.056 & 0.066 & 2.774 & 0.707 & 124.488 & & & & & \\
\multirow{1}{*} SG-HVS50
& 319.051060 & 2.739 & $-$66.662 & $-$303.651 & 2.293 & 555.367 & 0.47 & 0.46 & 0.54 & 0.57 & 0.55 \\
2689713751472944384 & 0.499367 & 0.104 & 0.086 & 4.087 & 0.551 & 169.175 & & & & & \\
\multirow{1}{*} SG-HVS51
& 338.674800 & $-$17.371 & $-$48.307 & $-$167.527 & 3.001 & 574.382 & 0.55 & 0.53 & 0.64 & 0.71 & 0.66 \\
2609260664602088704 & $-$8.696933 & 0.056 & 0.058 & 2.838 & 0.528 & 124.625 & & & & & \\
\multirow{1}{*} SG-HVS52
& 341.939260 & 36.388 & $-$5.165 & $-$91.327 & 3.357 & 520.554 & 0.29 & 0.28 & 0.52 & 0.66 & 0.55 \\
1888115422016149248 & 31.152207 & 0.020 & 0.025 & 2.790 & 0.292 & 48.674 & & & & & \\
\cline{1-12}
\end{tabular}}
\begin{tablenotes}
\footnotesize
\item[a]: Unbound probability calculated from the Galactic escape velocity curve under the Milky Way's potential {\tt MilkyWayPotential} implemented in {\tt Gala};
\item[b]: Unbound probability calculated from the Galactic escape velocity curve under the Milky Way's potential adopted from \citet{Monari2018};
\item[c]: Unbound probability calculated from the Galactic escape velocity curve under the Milky Way's potential adopted from \citet{Deason2019};
\item[d]: Unbound probability calculated from the Galactic escape velocity curve under the Milky Way's potential {\tt BovyMWPotential2014} implemented in {\tt Gala};
\item[e]: Unbound probability calculated from the Galactic escape velocity curve derived by \citet{Williams2017}.
\end{tablenotes}
\end{threeparttable}
\end{table*}

\begin{table*}[!b]
\centering
\begin{threeparttable}
\tablename{ 3: 8 HiVel pass through the Galactic plane within $1$\,kpc of the GC.}
\label{tbl:table2}
\resizebox{\textwidth}{!}{
\begin{tabular}{lllllllll}
\toprule
\multirow{1}{*} Notation & $\rm{ra}$ & $\mu_{\alpha}\rm{cos}\,\delta$ & $\mu_{\delta}$ & $v_{\rm{los}}$ & $d$ & $V_{\rm{GSR}}$  & Closest distance & Backward time \\ 
Gaia ID & $\rm{dec}$ & $\sigma_{\mu_{\alpha}\rm{cos}\,\delta}$ & $\sigma_{\mu_{\delta}}$  & $\sigma_{v_{\rm{los}}}$ & $\sigma_{d}$ & $\sigma_{V_{\rm{GSR}}}$  & $\sigma_{\rm{Closest\,distance}}$ & $\sigma_{\rm{Backward\,time}}$ \\
&  $\rm{(deg)}$ & $\rm{(mas\,yr^{-1})}$ & $\rm{(mas\,yr^{-1})}$ & $\rm{(km\,s^{-1})}$ & $\rm{(kpc)}$ & $\rm{(km\,s^{-1})}$ &  $\rm{(kpc)}$ & $\rm{(Myr)}$\\
\cline{1-9}
\multirow{1}{*} LG-HiVel4268
& 193.531234 & $-$40.214 & $-$0.964 & 107.820 & 2.804 & 441.795  & 0.64 & 16.3 \\
3934608033650104320 & 16.571303 & 0.049 & 0.050 & 14.730 & 0.344 & 57.339  & 0.42 & 1.2 \\
\multirow{1}{*} LG-HiVel6586
& 224.735724 & 5.904 & $-$25.086 & $-$196.640 & 3.998 & 431.806  & 0.04 & 878.2 \\
1268258491950103936 & 26.419553 & 0.043 & 0.046 & 16.890 & 0.777 & 80.058  & 3.80 & 390.0 \\
\multirow{1}{*} LG-HiVel6589
& 224.766129 & $-$27.429 & 11.510 & $-$94.390 & 3.226 & 430.846  & 0.37 & 14.7 \\
1262339339821422464 & 22.203286 & 0.049 & 0.048 & 16.310 & 0.504 & 60.393  & 0.38 & 1.5 \\
\multirow{1}{*} SG-HiVel7673
& 237.236130 & 5.080 & $-$32.384 & $-$142.706 & 3.573 & 460.369  & 0.83 & 1114.2 \\
1203663558210616448 & 18.710947 & 0.040 & 0.040 & 2.746 & 0.679 & 95.586  & 3.03 & 484.0 \\
\multirow{1}{*} SG-HiVel7983
& 241.245110 & 6.620 & $-$49.239 & 3.726 & 2.270 & 444.759  & 0.16 & 858.8 \\
1199253206614529280 & 16.786948 & 0.054 & 0.061 & 1.629 & 0.391 & 81.117  & 3.43 & 412.0 \\
\multirow{1}{*} LG-HiVel11799
& 324.655200 & $-$19.278 & $-$18.812 & $-$291.440 & 3.851 & 436.822  & 0.06 & 905.0 \\
1767040049125735168 & 12.585961 & 0.039 & 0.036 & 9.570 & 0.582 & 69.724  & 3.96 & 404.0 \\
\multirow{1}{*} LG-HiVel11997
& 330.291232 & 21.953 & 9.201 & $-$129.920 & 3.420 & 418.858  & 0.11 & 16.5 \\
1782324257184332032 & 21.134740 & 0.049 & 0.039 & 14.720 & 0.509 & 53.527  & 0.37 & 1.3 \\
\multirow{1}{*} SG-HiVel12087
& 333.417060 & $-$37.310 & $-$29.114 & $-$346.033 & 2.162 & 490.319  & 0.70 & 2853.4 \\
1782213107726332288 & 22.600503 & 0.052 & 0.049 & 2.919 & 0.269 & 56.338  & 3.91 & 1400.0 \\
\cline{1-9}
\end{tabular}}
\end{threeparttable}
\end{table*}

\begin{table*}[!b]
\centering
\begin{threeparttable}
\tablename{ 4: Extreme velocity stars probably originating from the Sgr dSph}
\label{tbl:table2}
\resizebox{\textwidth}{!}{
\begin{tabular}{lllllllll}
\toprule
\multirow{1}{*} Notation & $\rm{ra}$ & $\mu_{\alpha}\rm{cos}\,\delta$ & $\mu_{\delta}$ & $v_{\rm{los}}$ & $d$ & $V_{\rm{GSR}}$   & Closest distance & Backward time \\ 
Gaia ID & $\rm{dec}$ & $\sigma_{\mu_{\alpha}\rm{cos}\,\delta}$ & $\sigma_{\mu_{\delta}}$  & $\sigma_{v_{\rm{los}}}$ & $\sigma_{d}$ & $\sigma_{V_{\rm{GSR}}}$  & $\sigma_{\rm{Closest\,distance}}$ & $\sigma_{\rm{Backward\,time}}$\\
&  $\rm{(deg)}$ & $\rm{(mas\,yr^{-1})}$ & $\rm{(mas\,yr^{-1})}$ & $\rm{(km\,s^{-1})}$ & $\rm{(kpc)}$ & $\rm{(km\,s^{-1})}$ &  $\rm{(kpc)}$ & $\rm{(Myr)}$\\
\cline{1-9}
\multirow{1}{*} LG-HiVel62
& 2.375881 & 44.922 & 45.648 & $-$312.740 & 1.475 & 442.620  & 1.53 &  43.7 \\
383206057417413248 & 40.684454 & 0.023 & 0.021 & 12.970 & 0.068 & 19.154  & 0.67 & 2.1 \\
\multirow{1}{*} SG-HiVel1348
& 110.785890 & 40.379 & $-$31.264 & 383.674 & 1.573 & 421.659  & 3.80 &  44.0 \\
892923292818436224 & 32.062197 & 0.048 & 0.041 & 2.568 & 0.120 & 12.508  & 1.04 &  2.0\\
\multirow{1}{*} SG-HiVel1713
& 131.827190 & 40.839 & $-$29.768 & 457.035 & 1.411 & 436.414  & 3.70 &  53.4 \\
610169515364401024 & 16.154381 & 0.051 & 0.039 & 2.466 & 0.104 & 11.433  & 1.69 &  3.3\\
\multirow{1}{*} LG-HiVel2761
& 162.632256 & 3.450 & $-$29.152 & 503.080 & 1.634 & 426.484  & 3.46 &  56.4 \\
3982465926515084544 & 16.764019 & 0.034 & 0.026 & 12.330 & 0.079 & 11.959  & 1.31 &  3.6\\
\multirow{1}{*} RG-HiVel3317
& 175.273875 & $-$8.736 & 0.023 & 507.865 & 2.313 & 416.561  & 0.76 &  53.5 \\
3793871060689209984 & $-$1.545361 & 0.027 & 0.016 & 1.259 & 0.108 & 1.140  & 0.51 &  2.7\\
\multirow{1}{*} SG-HiVel5496
& 211.197860 & $-$33.479 & $-$3.457 & 248.658 & 3.143 & 446.331  & 4.71 &  40.0 \\
1231404094841937536 & 15.718307 & 0.051 & 0.037 & 2.736 & 0.538 & 61.237 & 0.90 &  3.3 \\
\multirow{1}{*} SG-HiVel6195
& 220.857380 & $-$46.968 & $-$1.443 & 193.119 & 2.453 & 472.287  & 2.93 &  40.5 \\
1186023710910901760 & 14.893445 & 0.052 & 0.041 & 3.011 & 0.267 & 46.884  & 0.83 &  2.8\\
\multirow{1}{*} LG-HiVel6293
& 221.804648 & $-$47.015 & $-$14.298 & 250.570 & 2.214 & 436.291  & 3.08 &  50.5 \\
1186213827638356096 & 15.009634 & 0.045 & 0.044 & 16.890 & 0.198 & 33.787 & 1.30 &  3.8\\
\multirow{1}{*} LG-HiVel6479
& 223.752833 & $-$41.880 & $-$7.281 & 184.430 & 2.529 & 438.804  & 1.98 &  45.2 \\
1282270187101294592 & 30.321404 & 0.050 & 0.056 & 13.280 & 0.374 & 54.743  & 1.44 &  3.2\\
\multirow{1}{*} SG-HiVel6578
& 224.688950 & $-$49.925 & $-$9.999 & 264.688 & 2.086 & 436.114  & 2.81 &  51.5 \\
1155161209793409408 & 4.451183 & 0.052 & 0.052 & 2.334 & 0.240 & 40.068 &  1.62 &  4.7\\
\multirow{1}{*} SG-HVS43
& 227.350790 & $-$42.685 & 11.386 & 38.323 & 3.395 & 637.638  & 4.94 &  29.8 \\
1263758598878844288 & 23.835252 & 0.059 & 0.072 & 2.909 & 0.780 & 151.478 & 2.09 &  3.0 \\
\multirow{1}{*} SG-HVS47
& 236.220050 & $-$39.882 & 13.278 & 87.606 & 2.953 & 556.258  & 4.17 &  36.1 \\
4429852530637477760 & 7.130787 & 0.064 & 0.051 & 4.646 & 0.545 & 95.418 & 1.47 &  4.0\\
\multirow{1}{*} SG-HiVel8799
& 251.449790 & $-$54.018 & 2.255 & 93.628 & 2.152 & 491.882  & 4.74 &  49.3 \\
4565596694311458816 & 21.149876 & 0.047 & 0.052 & 3.257 & 0.319 & 67.504  & 1.03 &  5.0\\
\multirow{1}{*} SG-HiVel12019
& 330.868640 & 4.706 & 52.478 & $-$317.809 & 1.722 & 452.259  & 2.28 &  41.9 \\
1962742265496273920 & 44.748228 & 0.084 & 0.076 & 2.985 & 0.271 & 66.563  & 1.64 &  2.8\\
\multirow{1}{*} RG-HiVel12112
& 334.496042 & $-$19.547 & 3.082 & $-$491.044 & 0.618 & 422.800  & 2.83 &  50.0 \\
2626144254057562112 & $-$5.197000 & 0.019 & 0.019 & 0.917 & 0.007 & 0.788  & 0.39 &  2.5\\
\cline{1-9}
\end{tabular}}
\end{threeparttable}
\end{table*}

\section{Comparisons with Other Work}
\citet{Du2018} presented a sample of 24 HiVel candidates with $V_{\rm GSR} \ge 0.85V_{\rm esc}$ from LAMOST\,DR5 and {\it Gaia}\,DR2.
From cross-matching our HiVel sample with theirs, 9 stars are found in common. 
The other 15 stars are not in our sample because they are excluded by our cuts:  1 star with relative parallax uncertainty larger than 20\%, 9 stars with parallax smaller than 0.2\,mas, 1 star with $v_{\rm los}$ variation larger than 15\,km\,s$^{-1}$, 3 stars with RUWE$\geq 1.4$, and 1 star with $V_{\rm GSR} < 300$\,km\,s$^{-1}$.

For stars in common, the distances adopted by \citet{Du2018} are systematically larger, by over 20\%, than those used in the current work.
The main reason is that in \citet{Du2018} the distances were determined from {\it Gaia} DR2 parallaxes with the official zero-point corrections. As shown by \citet{Schonrich2019} and \citet{Zinn2019}, the official {\it Gaia} DR2 parallax zero-point of  $0.029$\,mas, found by quasars \citep{Lindegren2018}, is smaller than the value of about $0.05$\,mas found for bright stars.
Given the bright nature of the LAMOST targets, the zero-point offset adopted by \citet{Du2018} is not sufficient, and thus the derived distances and velocities are overestimated.
We also note that all 9 of the stars in common are not HVS candidates.

Most recently, \citet[][hereafter Li21]{Li2021} present 591 HiVel candidates with total velocity ($V_{\rm GSR}$) greater than 445 km\,s$^{-1}$, selected from LAMOST\,DR7 and {\it Gaia}\,DR2.
Among those 591 HiVel candidates, 249 stars are found in our sample 
(including 12 classified as HVS candidates\footnote{One (LG-HVS35) of the 12 HVS candidates was actually first discovered by \citet{Bromley2018}.}; see the blue filled circles in Fig.\,3).
The remaining 342 stars are excluded by our various cuts (i.e., 125 stars either not included in LAMOST DR8 or without stellar parameters determined, 6 stars with spectral S/N smaller than 10, 5 stars with bad spectra excluded by visual checks, 159 stars with parallax smaller than 0.2\,mas, 1 star with relative parallax uncertainty larger than 20\%, 1 star with radial velocity uncertainty larger than 30\,km\,s$^{-1}$, 9 stars with $v_{\rm los}$ variation larger than 15\,km\,s$^{-1}$, 5 stars classified as variable stars, 14 stars with RUWE$\geq 1.4$, and 17 stars with total velocity $V_{\rm GSR} < 300$\,km\,s$^{-1}$).
For the stars in common, the total velocities yielded by Li21 are systematically larger than those derived in the current work, mainly due to their distances estimated directly by inverting {\it Gaia} DR2 parallax (without zero-point corrections).

In addition to the above searches, several other systematical searches purely based on {\it Gaia} DR2 have also reported hundreds of HiVel/HVS candidates \citep{Bromley2018,Hattori2018,Du2019,Li2020,Marchetti2019,Marchetti2021,Reggiani2022,QuispeHuaynasi2022,Liao2023}.
Comparing our sample with those candidates, we find 3 (including one star, LG-HVS35, which is also included in Li21; see above), 3, 1 and 5 candidates in common with \citet{Bromley2018}, \citet{Hattori2018}, \citet{Du2019} and \citet{Li2020}, respectively. It is interesting to note that two of our Hivel stars with $V_{\rm GSR}$ no more than 380\,km\,s$^{-1}$ were recently observed by high-resolution spectroscopy by \citet{Reggiani2022}. The metallicity and [$\alpha$/Fe] used in this work are consistent with those measured from high-resolution spectroscopy of \citet{Reggiani2022}.

\section{Probable Origins of Extreme Velocity Stars}

As shown in Figs.\,1 and 3, all the HVS candidates discovered in the present work are old, metal-poor, [$\alpha$/Fe]-enhanced, late-type halo stars.
In contrast, the known HVSs in the literature are mostly early-type, massive young stars, largely found from extreme radial velocity only.
Moreover, Fig.\,1 clearly shows that the ejection ratios of these late-type HVS candidates are systematically lower than those of the early-type stars. 
These properties suggest that the previously known HVSs (early-type) 
and our new HVS candidates (late-type) may have different ejection mechanisms. One of the explanation is that these HVS candidates are actually extreme bound halo stars as proposed by \citet{Hattori2018}. These stars are therefore can be used to constrain the Galactic potential. To constrain other possible origins of the late-type HVS candidates, we perform backward orbital integration 
in a model Galactic potential with the package {\tt Gala} for the 547 extreme velocity stars with $V_{\rm GSR} \ge 0.8V_{\rm esc}$ \citep[$V_{\rm esc}$ is taken from][]{Williams2017} in our sample (with only 15 of them classified as disk stars).
In Subsection \ref{subsec:orgin_bigPicture}, we will also explore the low-velocity portion of our HiVel sample
(namely, 300\,km$\,$s$^{-1} \le V_{\rm GSR} < 0.8V_{\rm esc}$).  
The Galactic potential of the MW adopted here is  {\tt MilkyWayPotential} in {\tt Gala}, consisting of four components, a nucleus, a bulge, a disk, and a dark matter halo. 
The details of the four components are described in \citet{Bovy2015}.
In the orbit integration, a time step of 0.1\,Myr is adopted.
Our explorations are described in the following three sections.

\subsection{Origination from the Galactic Center \label{subsec:orgin_GC}}
 
We first check whether any of our extreme velocity stars could have been ejected by the Galactic central SMBH via the Hills mechanism.  We select those stars whose backward-integrated orbits have their closest distances to the GC smaller than 1\,kpc as possible GC-originating candidates.
We note that only the candidates during the last pericentric passage through our Galaxy are considered.
In total, 8 extreme velocity stars are found to have an encounter with the GC within 1\,kpc, but none of them are HVS candidates, especially those with highest values of $v_{\rm GSR}$ (i.e. greater than 700\,km\,s$^{-1}$) that are expected from Hills mechanism \citep{Generozov2022}.
The lack of detections of HVSs with possible GC origins is similar to recent systematic searching efforts with {\it Gaia} data \citep{Marchetti2022}.
The reason is still unclear, but may due to the small volume covered by {\it Gaia} observations due to the requirement of well-measured parallaxes.
Table\,3 presents the information for those 8 stars, including orbital parameters (closest distance and backward time) and their uncertainties yielded from 2000 Monte Carlo trajectory calculations based on observational errors. 
The backward-integrated orbits are shown in the left panel of Fig.\,7.
By considering the observational uncertainties, we perform Monte Carlo simulations to infer the distribution in the $X$-$Y$ plane of each star during its intersection of the Galactic plane ($Z = 0$), under the assumption of GC origin. The resulted contours of those candidates, corresponding to the 90\% confidence level, are shown in the right panel of Fig.\,7. We note that the contour of one extreme velocity star, LG-HiVel\,11799, is very close to the GC.

Here, only 8 extreme velocity stars with $V_{\rm GSR}$ no more than 500\,km\,s$^{-1}$ are found to have had an encounter to the GC within 1\,kpc.
However, normal halo stars can naturally pass through the GC according to their velocity dispersion.
We thus perform a Monte Carlo simulation to evaluate the possibility of normal passages through the GC for halo stars with high velocity.
Doing so, one million halo stars are assumed uniformly distributed in a (5 kpc)$^{3}$ cubic space of $X$, $Y$ and $Z$, with 3D velocities following the distributions of local halo stars constructed by \citet{Anguiano2020}.
Similar to the above analysis, we perform backward orbital calculations for 16,041 extremely high-velocity simulated stars with $V_{\rm GSR} \ge 0.8 V_{\rm esc}$.
In total, 1342 of them are found to have a close encounter to the GC within 1\,kpc.
The derived fraction of 8.37\% (1342/16041) is even higher than the observed fraction of 1.46\% (8/547), indicating that the 8 extreme velocity stars with close encounters to the GC are probably normal halo stars passing through the GC by chance rather than having been ejected by interaction with the SMBH in the GC.


\subsection{Origination from the Sgr dSph \label{subsec:orgin_SgrGal}}
To further explore other origins of these extreme velocity stars, we integrate the backward trajectories for the following 
star clusters and dwarf galaxies:
1743 open clusters  with full 6D information and [Fe/H] from \citet{Dias2021};
147 Globular clusters with [Fe/H] from \citet{Harris2010}, radial velocities from \citet{Vasiliev2019}, proper motions from \citet{Vasiliev2021}, and distances from \citet{Baumgardt2021}; 
39 dwarf galaxies with full 6D information from \citet{Fritz2018}; 
and for the LMC and SMC \citep{Patel2020}. 

Possible links in the trajectories between the extreme velocity stars and the above Galactic subsystems are investigated by sorting the ratios of the closest orbital distances to the subsystems radii
\footnote{Here, the radius of a Galactic subsystem we adopt is as follows:
	tidal radius for open and globular clusters, normal radius for LMC and SMC, 
	and two times the half-light radius for dwarf galaxies.}.
For open or globular clusters, a further cut on metallicity difference (between those extreme velocity stars and clusters) smaller than 0.2\,dex is applied.
As in the above analysis (Subsection \ref{subsec:orgin_GC}), we only consider the last encounter between a specific star and a specific subsystem.
Interestingly, 15 extreme velocity stars (including 2 HVS candidates) are found to have close encounters with the Sgr dSph 
within two times its half-light radius \citep[$r_h = 2.59$\,kpc; ][]{McConnachie2012}.
Similar to previous investigations \citep{Huang2021,Bhat2022}, the uncertainties of the orbital parameters for each extreme velocity star are derived from the probability distribution functions (PDF) generated by 2000 Monte Carlo trajectory calculations, assuming that the measurement errors of both extreme velocity star and cluster/dwarf galaxy are
normally distributed, except for the distance of extreme velocity star (for which the
posterior PDF derived in Section 3.2 is used directly).
This information, together with their orbital parameters (closest distance and backward time), as well as their uncertainties, of the 15 extreme velocity stars are listed in Table 4.


As shown in Figs.\,8 and 9, the backward-integrated orbital analysis suggests that almost all the 15 extreme velocity stars intersect the trajectory of the Sgr dSph 
roughly at its latest pericenter (backward time of $39.7 \pm 1.9$ Myr) orbiting the MW.
By assuming their Sgr dSph origin, the stars have ejection velocities ranging from 500 to 900 km\,s$^{-1}$ when they left the Sgr dSph (see Fig.\,9).
Note that one of the candidates, J1443$+$1453, was already reported by \citet{Huang2021} to be the first HVS candidate originating from the Sgr dSph.  With new astrometric measurements from {\it Gaia} EDR3, 
it has a smaller total velocity than that derived from DR2, and thus is no longer a HVS.
Remarkably, 2 new HVS candidates, SG-HVS43 and SG-HVS47, are found to have had close encounters with the Sgr dSph around its pericenter, 
29.8 and 36.1 Myr ago, respectively.

Fig.\,10 shows these 15 extreme velocity stars\footnote{Note that three of them do not have [$\alpha$/Fe] measurements.} 
in the [$\alpha$/Fe] vs. [Fe/H] plane.
Half of them exhibit significantly lower [$\alpha$/Fe] abundance ratios compared to Galactic field stars \citep{Venn2004}, consistent with the distribution of the Sgr stream member stars.
There seem 5 other stars (including one HVS candidate) apparently with high  [$\alpha$/Fe] abundances, different from that of the Sgr stream member stars.
However, we note that their [$\alpha$/Fe] abundances were measured from low-resolution spectra, either by LAMOST or SEGUE, and thus the typical error to the abundances is 0.10 dex \citep{Lee2011, Xiang2019}.
High-resolution follow-up observations are required to clarify the chemical nature of those stars (including the three stars without [$\alpha$/Fe] measurements), and would be useful to test whether they are ejected from the Sgr dSph or not.

Finally, we note the independent work by \citet{LiHefan2022}, following the idea of \citet{Huang2021}, who report the discovery of 60 high-velocity star candidates originating in the Sgr dSph, using data from  {\it Gaia} EDR3 and several massive spectroscopic surveys.
The number of high-velocity star candidates in their work is larger than ours, mainly due to our cut on requiring reliable parallax measurements ($\varpi \ge 0.2$\,mas). We also note that 9 high-velocity RR Lyrae stars were recently identified by \citet{Prudil2022}, and 2 of them are tentatively linked to the Sgr dSph.

\begin{figure}[!b]
\begin{center}
\includegraphics[scale=0.32,angle=0]{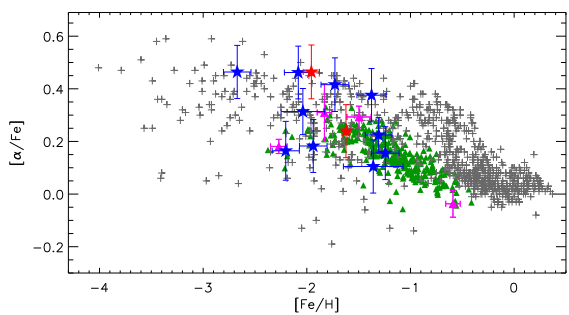}
\caption{Distributions of the extreme velocity star (blue stars) and HVS candidate (red stars) originating from the Sgr dSph galaxy,  Sgr stream member stars (green triangles), Sgr dSph-associated globular clusters (magenta squares), and field stars of the MW (gray plus symbols) in the [$\alpha$/Fe]--[Fe/H] plane.
The four Sgr dSph-associated globular clusters are M 54, Terzan 7, Terzan 8, and Arp 2, respectively. 
Their elemental-abundance ratios are measured from high-resolution spectroscopy \citep{Sbordone2005,Carretta2010,Carretta2014,Mottini2008}.}
\end{center}
\end{figure}

\begin{figure*}[!t]
\begin{center}
\includegraphics[scale=0.52,angle=0]{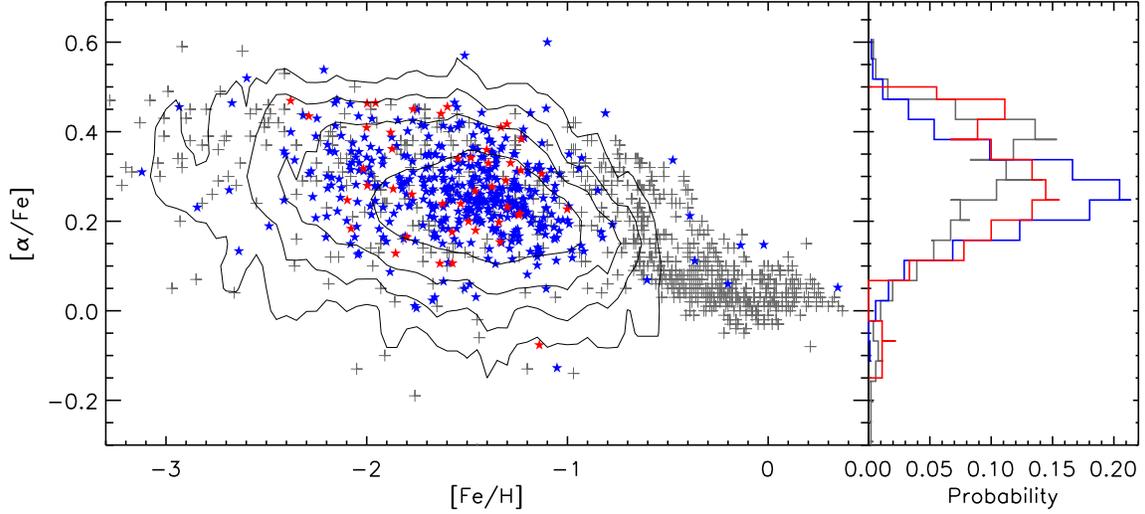}
\caption{\text{{\it Left panel}:} Similar to Fig.\,10, but for all extreme velocity stars (blue stars) with $V_{\rm GSR} \ge 0.8V_{\rm esc}$, HVS candidates (red stars), and field stars of the MW (gray pluses). The contours show the number density for the halo HiVels with $V_{\rm GSR} < 0.8V_{\rm esc}$. {\it Right panel}: The normalized distributions of [$\alpha$/Fe] for field halo stars (gray), HVS candidates (red) and extreme velocity stars (blue) with [Fe/H] between $-3.3$ and $-1.0$.}
\end{center}
\end{figure*}

\begin{figure}[!b]
\begin{center}
\includegraphics[scale=0.42,angle=0]{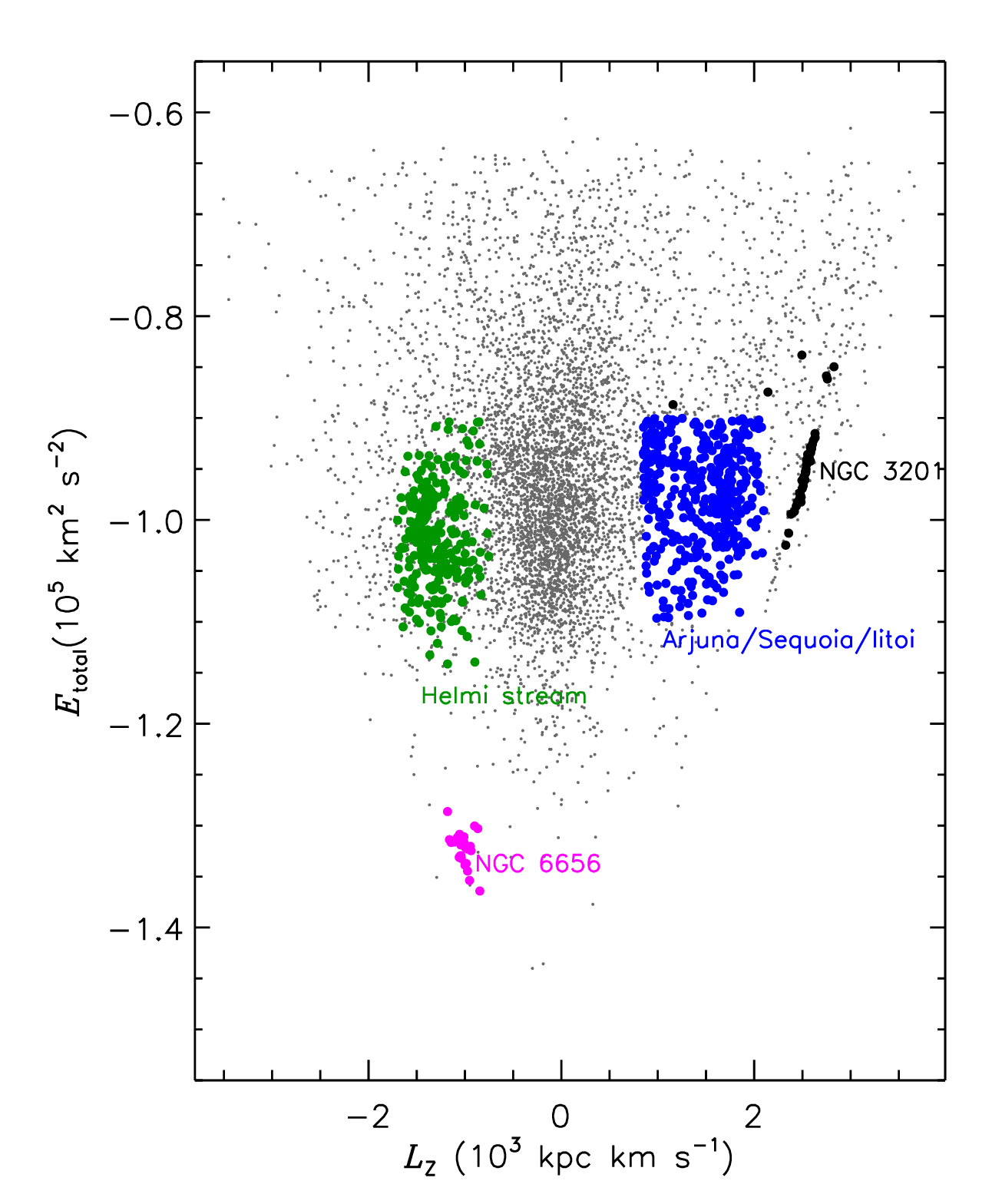}
\caption{Distribution of the 6434 halo HiVel stars (gray dots) with $V_{\rm GSR} < 0.8V_{\rm esc}$ 
	in the $E_{\rm total}$--$L_{\rm Z}$ diagram.
	We identify and denote some HiVels that belong to certain known Galactic subsystems (see text).
}
\end{center}
\end{figure}

In summary, the above orbital analysis show 15 extreme velocity stars (including 2 HVS candidates) encounter the Sgr dSph within its $2 r_h$ radius, 
when it passes by the latest pericenter of its orbit about the MW. Further analysis in terms of the [$\alpha$/Fe]--[Fe/H] diagram 
indicates that at least half of the stars have chemical abundances similar to the Sgr stream member stars and the Sgr dSph-associated globular clusters.
As discussed in \citet{Huang2021}, such stars are probably stripped from the tidally disrupted Sgr dSph 
during its latest pericentric passage, in agreement with the theoretical predictions by \citet{Abadi2009}. 
Moreover, the properties of this stellar population, together with further numerical simulations, 
can provide vital constraints on the theory proposed by \citet{Abadi2009}. 
Finally, as mentioned in \citet{Huang2021}, a second possibility is that 
these extreme velocity stars are ejected from the Sgr dSph via the Hills mechanism \citep{Hills1988}, 
provided a (central) massive/intermediate-mass black hole resides in the Sgr dSph.

\subsection{A General Picture: Tidal Ejection from Galactic Subsystems \label{subsec:orgin_bigPicture}}
The above finding, that 15 extreme velocity stars (including 2 HVS candidates) probably originated from the Sgr dSph, 
encourage us to consider a more general picture: 
Tidal ejections of stars from subsystems of our Galaxy can be an important channel to produce extreme velocity stars or even HVSs.

Here we explore a few more preliminary results for our sample.
First, we consider the relatively high-velocity portion of our sample, namely the extreme velocity stars with $V_{\rm GSR} \ge 0.8V_{\rm esc}$.
As can be seen from inspection of  Fig.\,11, the peak values (+0.25 to +0.30) of the [$\alpha$/Fe] distributions of $V_{\rm GSR} \ge 0.8V_{\rm esc}$ sources (including HVS candidates) are lower than the value (+0.40) found for the Galactic field halo stars.
 The lower [$\alpha$/Fe] abundance ratios of those stars are in-line with the chemical-evolution history of dwarf galaxies of the MW \citep[e.g.,][]{Tolstoy2009,Kirby2011}. Certainly, comprehensive numerical simulations (by considering dynamical effects), together with follow-up observations with high-resolution spectroscopy, are required to confirm this general picture, and to identify more extreme velocity stars or HVSs (especially the old metal-poor ones) that originated from Galactic subsystems.

We also comment on the sources with relatively lower kinetic energy, i.e., the portion with 300\,km$\,$s$^{-1} \le V_{\rm GSR} < 0.8V_{\rm esc}$ in our sample.
Their distribution in the [$\alpha$/Fe]--[Fe/H] diagram is also presented in Fig.\,11 (see the contours), and is broadly consistent with the extreme velocity star portion.
Furthermore, compared with the aforementioned extreme velocity stars, the lower kinetic energy stars have an additional merit that they better reflect the phase-space information of their progenitors.
Thus, following the strategy of \citet{Helmi1999}, we plot them in the diagram of the total energy ($E_{\rm total}$) vs. angular momentum ($L_{\rm Z}$), as shown in Fig.\,12.
We have identified the HiVels from our sample that belong to some well-known Galactic substructures and star clusters. The member stars of the Helmi stream are identified according to the criteria given by \citet{Koppelman2019}; those of Arjuna/Sequoia/I'itoi are from \citet{Naidu2020}.

For identifying the members of the globular cluster NGC\,3201 and its tidal stream, as well as  NGC\,6656,
we take the following criteria:
HiVel stars with line-of-sight velocity and metallicity similar to the given globular cluster, namely
$|v_{\rm los} - v_{\rm los}^{\rm cluster}| \le 30$\,km\,s$^{-1}$
and $|[{\rm Fe/H}] - [{\rm Fe/H}]^{\rm cluster}| \le$\,0.3\,dex.
The information for the clusters are taken from \citet{Harris2010}.
These substructures are also shown in Fig.\,12.
We can see that  
some of our HiVel sources definitely belong to certain known substructures, 
either simply (once) being the members of globular clusters (e.g., NGC\,3201 and NGC\,6656), 
or likely being the accreted debris of certain small systems (e.g., the aforementioned streams). 
This suggests that we can identify the Galactic-subsystem origins for these lower-velocity HiVel halo stars in the future, as larger and more accurate data are acquired.

\section{Summary}
Based on data from massive large-scale spectroscopic surveys, including the spectra of 
RAVE\,DR5, SDSS\,DR12, LAMOST\,DR8, APOGEE\,DR16, and GALAH\,DR2, 
and the {\it Gaia} EDR3 astrometry, 
we have assembled a large sample of 12,784 HiVel stars with $V_{\rm GSR} \ge 300$\,km\,s$^{-1}$.
Importantly, 52 HVS candidates with total velocity marginally exceeding their local escape velocities are found in this sample, 40 of which are discovered for the first time.
More interestingly, all the candidates are late-type metal-poor stars.
The properties of these candidates are significantly different from 
the previous HVSs in the literature, which are generally massive and early-type, 
and were primarily found by their extreme radial velocities alone.

We perform orbital analyses for 547 extreme velocity stars with $\it{V}_{\rm{GSR}}/V_{\rm{esc}}\geq\rm{0.8}$ in our sample to investigate their possible origins. A total of 15 extreme velocity stars (including 2 HVS candidates) are found to intersect with the orbit of the Sgr dSph 
within its $2 r_h$ around its latest pericentric passage through the MW.
Analysis of the [$\alpha$/Fe]--[Fe/H] diagram 
indicates that at least half of the 15 stars have chemical abundances similar to the Sgr dSph stars.
We have thus sought evidence for the origins of additional HiVels in our sample.
After a preliminary analysis, we propose a general picture:
Star ejection from Galactic subsystems such as dwarf galaxy and globular clusters, 
either via tidal stripping or even the Hills mechanism, 
can be an important channel to produce extreme velocity stars or even HVSs, 
particularly the metal-poor late-type population discovered in the present study.

\section*{Acknowledgments}
{We thank an anonymous referee for helpful comments.
This work is supported by National Natural Science Foundation
of China grants 11903027, 11973001, 11833006, 11873083,
U1731108, 12090040, 12090044, and National Key R \& D Program of China No. 2019YFA0405500. Y.H. is supported
by the Yunnan University grant No. C176220100006. T.C.B. acknowledges partial support
for this work from grant PHY 14-30152; Physics Frontier
Center/JINA Center for the Evolution of the Elements (JINA-CEE),
awarded by the US National Science Foundation.
HWZ acknowledges the science research grants from the
China Manned Space Project with No. CMS-CSST-2021-B03.
Z.Y. acknowledge support from the French National Research Agency (ANR) funded project (ANR-18-CE31-0017) and the European Research Council (ERC) under the European Unions Horizon 2020 research and innovation programme (grant agreement No. 834148).

The Guoshoujing Telescope (the Large Sky Area Multi-Object Fiber Spectroscopic Telescope, LAMOST) is a National Major Scientific Project built by the Chinese Academy of Sciences. Funding for the project has been provided by the National Development and Reform Commission. LAMOST is operated and managed by the National Astronomical Observatories, Chinese Academy of Sciences. The LAMOST FELLOWSHIP is supported by Special fund for Advanced Users, budgeted and administrated by Center for Astronomical Mega-Science, Chinese Academy of Sciences (CAMS). 

The GALAH survey is based on observations made at theAnglo-Australian  Telescope,  under  programmes  A/2013B/13,A/2014A/25, A/2015A/19, A/2017A/18. We acknowledge the traditional owners of the land on which the AAT stands, the Gamilaraay people, and pay our respects to elders past and present.

Funding for the Sloan Digital Sky Survey IV has been provided by the Alfred P. Sloan Foundation, the U.S. Department of Energy Office of Science, and the Participating Institutions. SDSS acknowledges support and resources from the Center for High-Performance Computing at the University of Utah. The SDSS web site is www.sdss.org.

Funding for RAVE has been provided by: the Australian Astronomical Observatory; the Leibniz-Institut fuer Astrophysik Potsdam (AIP); the Australian National University; the Australian Research Council; the French National Research Agency; the German Research Foundation (SPP 1177 and SFB 881); the European Research Council (ERC-StG 240271 Galactica); the Istituto Nazionale di Astrofisica at Padova; The Johns Hopkins University; the National Science Foundation of the USA (AST-0908326); the W. M. Keck foundation; the Macquarie University; the Netherlands Research School for Astronomy; the Natural Sciences and Engineering Research Council of Canada; the Slovenian Research Agency; the Swiss National Science Foundation; the Science \& Technology Facilities Council of the UK; Opticon; Strasbourg Observatory; and the Universities of Groningen, Heidelberg and Sydney. The RAVE web site is at https://www.rave-survey.org.

This work has made use of data from the European Space Agency (ESA) mission {\it Gaia} (\url{https://www.cosmos.esa.int/gaia}), processed by the {\it Gaia}
\url{https://www.cosmos.esa.int/web/gaia/dpac/consortium}). Funding for the DPAC
has been provided by national institutions, in particular the institutions
participating in the {\it Gaia} Multilateral Agreement.

\bibliographystyle{apj}
\bibliography{HVS_Notes}

\begin{appendix}

\section{Information of 88 known HVSs or candidates}
In Table\,A1, we present proper motions, line-of-sight velocities, distances, stellar atmospheric parameters, [$\alpha$/Fe], radial velocity in the Galactic rest frame $V_{\rm rf}$, and total velocity $V_{\rm GSR}$ for 88 known HVSs or candidates, including the 52 found in this work.
For those HVS candidates discovered in this work, the first letter of their names represents the spectroscopic survey used to find this HVS candidates (i.e., L -- LAMOST and S -- SDSS) and the second letter is always G, indicating the astrometric information from {\it Gaia} EDR3.
For those HVSs found by \citet{Brown2005,Brown2006,Brown2007,Brown2009,Brown2012,Brown2014}, the distances and stellar atmospheric parameters are all taken from their final updates \citep{Brown2014}.
The distances of the three WD HVSs, found by Shen et al. (2018), are derived by the parallax measurements from {\it Gaia} EDR3 by the method mentioned in Section\,3.2.
For the remaining HVSs or HVS candidates, their information are taken from the references listed in Table\,A1.
The proper motions for all those HVSs or HVS candidates are taken from {\it Gaia} EDR3.
$V_{\rm rf}$ and $V_{\rm GSR}$ are those calculated by adopting the values of $U_{\odot}$, $v_{\phi, \odot}$, $W_{\odot}$ and $R_0$ described in Section\,3.1.

\begin{table*}[!b]
\centering
\begin{threeparttable}
\tablename{ A1: Basic Parameters for 88 HVSs or candidates.}
\label{tbl:table2}
\resizebox{\textwidth}{!}{
\begin{tabular}{lrrrrrrrrrrrrr}
\toprule
\multirow{1}{*} ID & Notation & $\rm{ra}$ & $\mu_{\alpha}\rm{cos}\,\delta$ & $\mu_{\delta}$ & $v_{\rm{los}}$ & $\it{d}$  & $T_{\rm eff}$ & ${\rm log} g$ & [Fe/H]  & [$\alpha$/Fe]& $\it{V}_{\rm{rf}}$ & $\it{V}_{\rm{GSR}}$ & Reference\\ 
&  Gaia ID & $\rm{dec}$ & $\sigma_{\mu_{\alpha}\rm{cos}\,\delta}$ & $\sigma_{\mu_{\delta}}$  & $\sigma_{v_{\rm{los}}}$ & $\sigma_{\it{d}}$ & $\sigma_{T_{\rm eff}}$ & $\sigma_{{\rm log} g}$ & $\sigma_{\rm{[Fe/H]}}$  & $\sigma_{[\alpha/\rm{Fe}]}$ & $\sigma_{\it{V}_{\rm{rf}}}$& $\sigma_{\it{V}_{\rm{GSR}}}$ &  \\
& & $(\rm{deg})$ & $(\rm{mas\,yr^{-1}})$ & $(\rm{mas\,yr^{-1}})$ & $(\rm{km\,s^{-1}})$ & $(\rm{kpc})$ & (K) & (dex) & (dex)  & (dex)& $(\rm{km\,s^{-1}})$ & $(\rm{km\,s^{-1}})$ &  \\
\cline{1-14}
\multicolumn{14}{c}{HVS candidates from this work} \\
\cline{1-14}
\multirow{1}{*}
1 & LG-HVS1 &  12.362361 & 11.129 & $-$59.488 & $-$345.37 & 2.33 & 5629 & 4.29 & $-$1.46 & -- & $-$231.24 & 535.01 & this work \\
& 2556679991337823488 & 7.128955 & 0.087 & 0.078 & 9.48 & 0.47 & 201 & 0.32 & 0.19 & -- & 9.48 & 117.95 & \\
\multirow{1}{*}
2 & LG-HVS2 &  12.509631 & 48.314 & 1.221 & $-$32.94 & 2.71 & 6065 & 4.21 & $-$1.34 & $+$0.15 & 120.38 & 517.59 & this work$^a$ \\
 & 2801887851883799936 & 21.420701 & 0.065 & 0.039 & 14.38 & 0.44 & 257 & 0.40 & 0.24 & 0.09 & 14.38 & 93.60 & \\
\multirow{1}{*}
3 & SG-HVS3 &  29.303224 & $-$15.182 & $-$38.759 & 217.00 & 2.46 & 5961 & 4.24 & $-$1.24 & $+$0.22 & 267.09 & 507.09 & this work \\
& 2510946771548268160 & 1.193651 & 0.071 & 0.056 & 2.26 & 0.43 & 19 & 0.06 & 0.03 & 0.01 & 2.26 & 61.54 & \\
\multirow{1}{*}
4 & LG-HVS4 &  35.653263 & 75.919 & 10.855 & 151.05 & 1.51 & 5290 & 4.41 & $-$1.53 & $+$0.24 & 244.44 & 511.54 & this work$^a$ \\
 & 87667257538859264 & 21.079511 & 0.089 & 0.067 & 9.47 & 0.18 & 195 & 0.31 & 0.18 & 0.05 & 9.47 & 53.38 & \\
\multirow{1}{*}
5 & LG-HVS5 &  37.488910 & 57.169 & $-$40.756 & $-$263.45 & 2.76 & 5874 & 3.87 & $-$2.02 & $+$0.32 & $-$230.70 & 716.09 & this work$^a$ \\
 & 2503400067332349440 & 3.130835 & 0.065 & 0.055 & 18.68 & 0.54 & 194 & 0.32 & 0.19 & 0.06 & 18.68 & 169.39 & \\
\multirow{1}{*}
6 & LG-HVS6 &  54.010210 & 32.181 & $-$26.060 & 67.55 & 3.58 & 5969 & 4.04 & $-$1.57 & $+$0.18 & 173.13 & 504.38 & this work \\
& 237369721329578112 & 41.872442 & 0.053 & 0.039 & 13.92 & 0.70 & 55 & 0.09 & 0.05 & 0.04 & 13.92 & 128.47 & \\
\multirow{1}{*}
7 & SG-HVS7 &  62.335793 & $-$21.928 & $-$86.078 & 16.31 & 1.52 & 5463 & 4.53 & $-$1.38 & $+$0.28 & 18.31 & 532.71 & this work \\
& 45249542048091392 & 15.531954 & 0.182 & 0.125 & 3.30 & 0.32 & 58 & 0.11 & 0.08 & 0.01 & 3.30 & 124.37 & \\
\multirow{1}{*}
8 & LG-HVS8 &  84.087702 & 75.773 & $-$54.534 & $-$118.04 & 1.60 & 6098 & 4.08 & $-$1.57 & -- & $-$31.81 & 492.41 & this work \\
& 263868123355334784 & 53.192993 & 0.049 & 0.040 & 14.12 & 0.13 & 208 & 0.33 & 0.19 & -- & 14.12 & 57.82 & \\
\multirow{1}{*}
9 & SG-HVS9 &  107.714100 & $-$39.266 & $-$70.707 & $-$43.47 & 1.71 & 4900 & 4.37 & $-$2.38 & $+$0.47 & $-$41.10 & 519.62 & this work \\
& 946735552249457152 & 39.286968 & 0.091 & 0.080 & 1.81 & 0.30 & 78 & 0.10 & 0.11 & 0.01 & 1.81 & 107.15 & \\
\multirow{1}{*}
10 & SG-HVS10 &  111.533120 & 46.016 & $-$46.600 & $-$118.50 & 2.56 & 5630 & 4.46 & $-$1.23 & $+$0.31 & $-$130.90 & 600.59 & this work \\
& 897267428899789440 & 36.689452 & 0.080 & 0.064 & 1.91 & 0.56 & 32 & 0.05 & 0.03 & 0.01 & 1.91 & 166.71 & \\
\multirow{1}{*}
11 & SG-HVS11 &  122.591610 & $-$0.332 & $-$59.917 & $-$33.97 & 2.80 & 6223 & 3.92 & $-$1.87 & $+$0.36 & $-$37.24 & 549.73 & this work \\
& 921561993012731648 & 40.652329 & 0.065 & 0.043 & 3.20 & 0.60 & 35 & 0.12 & 0.05 & 0.01 & 3.20 & 169.48 & \\
\multirow{1}{*}
12 & SG-HVS12 &  141.546960 & $-$17.158 & $-$33.579 & 223.55 & 3.77 & 6201 & 3.33 & $-$1.77 & $+$0.45 & 160.23 & 484.93 & this work \\
& 694458695225567872 & 27.340210 & 0.051 & 0.037 & 2.03 & 0.70 & 30 & 0.04 & 0.04 & 0.01 & 2.03 & 115.51 & \\
\multirow{1}{*}
13 & SG-HVS13 &  147.216280 & 11.123 & $-$128.292 & 75.05 & 1.18 & 4673 & 4.44 & $-$2.29 & $+$0.43 & 157.41 & 514.64 & this work \\
& 1050729011271507328 & 60.705308 & 0.039 & 0.051 & 2.64 & 0.07 & 58 & 0.26 & 0.10 & 0.01 & 2.64 & 39.61 & \\
\multirow{1}{*}
14 & SG-HVS14 &  149.660010 & 39.411 & $-$23.042 & 38.19 & 2.85 & 6163 & 3.97 & $-$1.23 & $+$0.38 & 25.01 & 578.54 & this work \\
& 803054228887217024 & 38.134845 & 0.050 & 0.045 & 3.03 & 0.47 & 28 & 0.13 & 0.03 & 0.01 & 3.03 & 92.67 & \\
\multirow{1}{*}
15 & LG-HVS15 &  154.757012 & $-$17.830 & $-$40.305 & $-$134.73 & 3.54 & 5998 & 3.75 & $-$1.60 & $+$0.46 & $-$111.87 & 507.82 & this work \\
& 809462835487742080 & 45.733809 & 0.044 & 0.055 & 19.40 & 0.72 & 354 & 0.57 & 0.34 & 0.14 & 19.40 & 145.16 & \\
\multirow{1}{*}
16 & SG-HVS16 &  166.340900 & 43.549 & $-$18.727 & 313.31 & 2.40 & 5585 & 4.24 & $-$1.40 & $+$0.36 & 120.02 & 599.51 & this work \\
& 3559325645434651648 & $-$17.070308 & 0.088 & 0.064 & 2.57 & 0.46 & 44 & 0.08 & 0.04 & 0.02 & 2.57 & 97.66 & \\
\multirow{1}{*}
17 & SG-HVS17 &  167.381140 & $-$24.318 & $-$36.315 & 439.25 & 3.75 & 6304 & 3.55 & $-$2.00 & $+$0.28 & 247.00 & 662.14 & this work \\
& 3559089525313289344 & $-$17.284739 & 0.055 & 0.045 & 3.21 & 0.79 & 31 & 0.10 & 0.05 & 0.01 & 3.21 & 152.83 & \\
\multirow{1}{*}
18 & LG-HVS18 &  169.502405 & $-$36.101 & 4.092 & 157.24 & 3.10 & 6296 & 4.28 & $-$1.13 & $+$0.31 & 125.14 & 537.97 & this work \\
& 3998883554967849216 & 28.818153 & 0.050 & 0.081 & 16.14 & 0.39 & 174 & 0.28 & 0.17 & 0.05 & 16.14 & 58.38 & \\
\multirow{1}{*}
19 & LG-HVS18 &  170.792361 & $-$35.643 & $-$35.437 & 34.18 & 3.36 & 6044 & 4.13 & $-$1.57 & $+$0.11 & 50.23 & 577.20 & this work \\
& 770479307125812864 & 40.248147 & 0.055 & 0.058 & 15.53 & 0.74 & 173 & 0.28 & 0.16 & 0.07 & 15.53 & 173.47 & \\
\multirow{1}{*}
20 & LG-HVS20 &  180.581777 & $-$36.995 & $-$32.226 & 26.33 & 3.19 & 6130 & 4.07 & $-$1.77 & $+$0.26 & 24.25 & 513.93 & this work \\
& 4026489543162246912 & 31.837635 & 0.074 & 0.055 & 19.69 & 0.71 & 309 & 0.48 & 0.29 & 0.11 & 19.69 & 162.69 & \\
\multirow{1}{*}
21 & SG-HVS21 &  189.653260 & $-$43.531 & $-$7.011 & $-$66.74 & 3.20 & 6209 & 3.65 & $-$1.28 & $+$0.33 & $-$49.55 & 534.39 & this work$^a$ \\
 & 1514459756956334720 & 32.344768 & 0.057 & 0.056 & 3.49 & 0.55 & 38 & 0.10 & 0.02 & 0.01 & 3.49 & 106.64 & \\
\multirow{1}{*}
22 & LG-HVS22 &  191.641073 & $-$23.127 & $-$33.598 & $-$88.91 & 3.92 & 6202 & 4.20 & $-$1.30 & $+$0.23 & $-$73.63 & 510.82 & this work \\
& 1513259266353444992 & 30.695884 & 0.049 & 0.062 & 11.83 & 0.86 & 58 & 0.10 & 0.06 & 0.04 & 11.83 & 164.60 & \\
\multirow{1}{*}
23 & SG-HVS23 &  191.669490 & 27.761 & $-$30.708 & 360.73 & 1.94 & 4930 & 4.70 & $-$1.33 & $+$0.41 & 313.45 & 510.89 & this work$^a$ \\
 & 3929069346904353152 & 13.430034 & 0.081 & 0.075 & 1.58 & 0.26 & 13 & 0.09 & 0.02 & 0.01 & 1.58 & 32.83 & \\
\multirow{1}{*}
24 & SG-HVS24 &  191.991830 & $-$41.337 & $-$51.052 & 81.99 & 2.42 & 5090 & 4.59 & $-$1.39 & $+$0.33 & 162.49 & 540.11 & this work \\
& 1567338780126678400 & 48.967882 & 0.062 & 0.080 & 1.48 & 0.56 & 26 & 0.04 & 0.06 & 0.01 & 1.48 & 165.63 & \\
\multirow{1}{*}
25 & SG-HVS25 &  192.498020 & $-$52.733 & $-$67.243 & $-$168.67 & 1.90 & 6129 & 4.06 & $-$2.00 & $+$0.46 & $-$121.96 & 536.10 & this work$^a$ \\
 & 1520968079814522496 & 38.947287 & 0.030 & 0.032 & 2.32 & 0.18 & 23 & 0.14 & 0.04 & 0.01 & 2.32 & 69.35 & \\
\multirow{1}{*}
26 & LG-HVS26 &  194.347116 & $-$7.334 & $-$51.745 & 175.76 & 2.88 & 6007 & 4.42 & $-$1.32 & -- & 159.90 & 526.63 & this work \\
& 3942455900972145152 & 20.419876 & 0.069 & 0.052 & 12.66 & 0.47 & 282 & 0.44 & 0.26 & -- & 12.66 & 107.15 & \\
\multirow{1}{*}
27 & LG-HVS27 &  194.385615 & $-$16.537 & $-$43.399 & $-$221.79 & 3.59 & 6151 & 3.97 & $-$2.10 & $+$0.25 & $-$177.83 & 581.30 & this work$^a$ \\
 & 1517369756214955008 & 37.110194 & 0.023 & 0.026 & 15.93 & 0.51 & 38 & 0.06 & 0.04 & 0.05 & 15.93 & 106.32 & \\
\multirow{1}{*}
28 & LG-HVS28 &  196.007951 & $-$5.770 & $-$38.240 & $-$154.57 & 3.85 & 6154 & 4.07 & $-$1.48 & $+$0.34 & $-$137.44 & 514.01 & this work \\
& 1461216666591966336 & 28.552378 & 0.037 & 0.051 & 11.84 & 0.62 & 47 & 0.08 & 0.04 & 0.03 & 11.84 & 106.64 & \\
\multirow{1}{*}
29 & LG-HVS29 &  200.292967 & $-$74.467 & $-$56.547 & $-$58.03 & 1.83 & 5475 & 4.51 & $-$1.55 & -- & $-$64.88 & 576.62 & this work \\
& 3938867537399911680 & 18.835966 & 0.088 & 0.051 & 10.99 & 0.26 & 251 & 0.40 & 0.23 & -- & 10.99 & 114.30 & \\
\multirow{1}{*}
30 & LG-HVS30 &  201.229653 & $-$31.015 & $-$40.112 & 380.20 & 2.55 & 5269 & 4.65 & $-$1.34 & $+$0.20 & 383.05 & 526.69 & this work \\
& 1442286920356579712 & 20.939786 & 0.085 & 0.057 & 7.65 & 0.57 & 284 & 0.45 & 0.27 & 0.09 & 7.65 & 94.60 & \\
\multirow{1}{*}
31 & LG-HVS31 &  203.794513 & $-$43.676 & $-$9.073 & 238.43 & 3.07 & 6341 & 4.12 & $-$1.16 & -- & 206.53 & 524.69 & this work \\
& 3725682648069934336 & 8.899663 & 0.049 & 0.028 & 14.73 & 0.43 & 318 & 0.49 & 0.30 & -- & 14.73 & 79.09 & \\
\multirow{1}{*}
32 & SG-HVS32 &  204.084080 & $-$45.700 & $-$6.408 & 33.26 & 3.56 & 6222 & 4.00 & $-$1.46 & $+$0.27 & 5.90 & 620.15 & this work \\
& 3726042154012422656 & 10.005512 & 0.068 & 0.048 & 2.93 & 0.81 & 20 & 0.08 & 0.05 & 0.01 & 2.93 & 171.28 & \\
\multirow{1}{*}
33 & LG-HVS33 &  206.252711 & 15.691 & $-$29.640 & $-$104.20 & 3.75 & 6348 & 4.16 & $-$1.31 & $+$0.29 & $-$5.82 & 566.40 & this work$^a$ \\
 & 1503968702337282048 & 46.596244 & 0.030 & 0.038 & 16.34 & 0.54 & 156 & 0.26 & 0.15 & 0.07 & 16.34 & 78.93 & \\
\multirow{1}{*}
34 & SG-HVS34 &  207.085230 & 16.371 & $-$54.602 & 216.56 & 2.03 & 5030 & 4.66 & $-$1.30 & $+$0.42 & 299.65 & 559.24 & this work$^a$ \\
 & 1500305198313028224 & 40.949577 & 0.048 & 0.060 & 1.56 & 0.28 & 30 & 0.09 & 0.04 & 0.01 & 1.56 & 58.23 & \\
\multirow{1}{*}
35 & LG-HVS35 &  214.976283 & $-$42.857 & $-$17.186 & $-$245.32 & 3.36 & 6238 & 3.97 & $-$1.60 & -- & $-$154.56 & 549.05 & this work$^b$ \\
 & 1484524973071001984 & 37.669366 & 0.023 & 0.029 & 14.68 & 0.35 & 214 & 0.34 & 0.20 & -- & 14.68 & 71.49 & \\
\multirow{1}{*}
36 & SG-HVS36 &  216.185130 & $-$27.628 & $-$20.628 & $-$338.22 & 4.15 & 5473 & 2.76 & $-$1.80 & $+$0.16 & $-$220.42 & 506.90 & this work \\
& 1506678826700636544 & 46.550514 & 0.020 & 0.024 & 1.40 & 0.39 & 43 & 0.87 & 0.03 & 0.05 & 1.40 & 58.07 & \\
\cline{1-14}
\end{tabular}}
\end{threeparttable}
\end{table*}

\begin{table*}[!b]
\centering
\begin{threeparttable}
\tablename{ A1: Basic Parameters for 88 HVSs or candidates.}
\label{tbl:table2}
\resizebox{\textwidth}{!}{
\begin{tabular}{lrrrrrrrrrrrrr}
\toprule
\multirow{1}{*} ID & Notation & $\rm{ra}$ & $\mu_{\alpha}\rm{cos}\,\delta$ & $\mu_{\delta}$ & $v_{\rm{los}}$ & $\it{d}$  & $T_{\rm eff}$ & ${\rm log} g$ & [Fe/H]  & [$\alpha$/Fe]& $\it{V}_{\rm{rf}}$ & $\it{V}_{\rm{GSR}}$ & Reference\\ 
&  Gaia ID & $\rm{dec}$ & $\sigma_{\mu_{\alpha}\rm{cos}\,\delta}$ & $\sigma_{\mu_{\delta}}$  & $\sigma_{v_{\rm{los}}}$ & $\sigma_{\it{d}}$ & $\sigma_{T_{\rm eff}}$ & $\sigma_{{\rm log} g}$ & $\sigma_{\rm{[Fe/H]}}$  & $\sigma_{[\alpha/\rm{Fe}]}$ & $\sigma_{\it{V}_{\rm{rf}}}$& $\sigma_{\it{V}_{\rm{GSR}}}$ &  \\
& & $(\rm{deg})$ & $(\rm{mas\,yr^{-1}})$ & $(\rm{mas\,yr^{-1}})$ & $(\rm{km\,s^{-1}})$ & $(\rm{kpc})$ & (K) & (dex) & (dex)  & (dex)& $(\rm{km\,s^{-1}})$ & $(\rm{km\,s^{-1}})$ &  \\
\cline{1-14}
\multirow{1}{*}
37 & SG-HVS37 &  217.396790 & $-$3.753 & $-$54.087 & $-$112.50 & 2.91 & 6408 & 3.75 & $-$1.88 & $+$0.40 & $-$48.10 & 582.99 & this work$^a$ \\
 & 1280443412952917632 & 26.915099 & 0.040 & 0.039 & 2.45 & 0.37 & 15 & 0.09 & 0.03 & 0.01 & 2.45 & 91.52 & \\
\multirow{1}{*}
38 & SG-HVS38 &  218.817840 & $-$47.200 & $-$36.613 & $-$237.30 & 2.68 & 6407 & 3.61 & $-$1.46 & $+$0.18 & $-$185.33 & 548.67 & this work \\
& 1242022529603565440 & 21.728278 & 0.026 & 0.031 & 1.96 & 0.26 & 37 & 0.13 & 0.06 & 0.01 & 1.96 & 67.94 & \\
\multirow{1}{*}
39 & LG-HVS39 &  224.735948 & 7.222 & $-$29.369 & $-$186.54 & 4.39 & 6004 & 4.25 & $-$1.14 & $-$0.08 & $-$51.06 & 580.33 & this work \\
& 1587301410160177920 & 46.728570 & 0.046 & 0.054 & 13.83 & 0.90 & 102 & 0.17 & 0.10 & 0.17 & 13.83 & 121.66 & \\
\multirow{1}{*}
40 & LG-HVS40 &  224.807592 & $-$16.170 & $-$65.173 & $-$66.60 & 2.28 & 5763 & 4.03 & $-$1.79 & -- & $-$10.13 & 525.83 & this work \\
& 1188512524199896064 & 17.866868 & 0.063 & 0.087 & 15.61 & 0.43 & 282 & 0.44 & 0.26 & -- & 15.61 & 131.91 & \\
\multirow{1}{*}
41 & LG-HVS41 &  226.584048 & $-$2.004 & $-$43.521 & $-$131.44 & 3.85 & 6114 & 4.00 & $-$1.49 & $+$0.20 & $-$49.17 & 642.76 & this work \\
& 1264622956753148416 & 24.765407 & 0.036 & 0.050 & 16.96 & 0.80 & 352 & 0.56 & 0.33 & 0.13 & 16.96 & 160.07 & \\
\multirow{1}{*}
42 & SG-HVS42 &  227.154450 & 3.842 & $-$30.310 & $-$211.83 & 4.09 & 5797 & 4.23 & $-$1.63 & $+$0.44 & $-$61.56 & 538.21 & this work \\
& 1592830751057210496 & 51.533092 & 0.051 & 0.058 & 2.18 & 0.91 & 22 & 0.07 & 0.03 & 0.01 & 2.18 & 122.54 & \\
\multirow{1}{*}
43 & SG-HVS43 &  227.350790 & $-$42.685 & 11.386 & 38.32 & 3.40 & 6212 & 4.23 & $-$1.62 & $+$0.24 & 119.92 & 637.64 & this work \\
& 1263758598878844288 & 23.835252 & 0.059 & 0.072 & 2.91 & 0.78 & 20 & 0.12 & 0.04 & 0.01 & 2.91 & 151.48 & \\
\multirow{1}{*}
44 & SG-HVS44 &  229.492430 & $-$38.172 & $-$15.989 & $-$269.74 & 3.89 & 6267 & 3.66 & $-$1.64 & $+$0.11 & $-$204.84 & 584.79 & this work \\
& 1208146095315810944 & 16.324317 & 0.069 & 0.056 & 4.13 & 0.91 & 46 & 0.06 & 0.06 & 0.01 & 4.13 & 163.72 & \\
\multirow{1}{*}
45 & SG-HVS45 &  232.301250 & $-$5.243 & $-$42.533 & $-$189.46 & 3.01 & 5691 & 4.29 & $-$1.55 & $+$0.34 & $-$31.61 & 513.10 & this work \\
& 1594746199095755520 & 50.496393 & 0.063 & 0.075 & 3.03 & 0.56 & 42 & 0.03 & 0.04 & 0.02 & 3.03 & 108.32 & \\
\multirow{1}{*}
46 & LG-HVS46 &  234.156763 & 5.004 & $-$38.522 & 16.63 & 4.50 & 5517 & 3.88 & $-$1.00 & $+$0.23 & 72.07 & 701.44 & this work \\
& 1164837294370596992 & 9.002340 & 0.043 & 0.044 & 8.13 & 0.90 & 57 & 0.09 & 0.05 & 0.03 & 8.13 & 158.66 & \\
\multirow{1}{*}
47 & SG-HVS47 &  236.220050 & $-$39.882 & 13.278 & 87.61 & 2.95 & 6030 & 3.60 & $-$1.96 & $+$0.46 & 143.06 & 556.26 & this work \\
& 4429852530637477760 & 7.130787 & 0.064 & 0.051 & 4.65 & 0.55 & 42 & 0.09 & 0.08 & 0.02 & 4.65 & 95.42 & \\
\multirow{1}{*}
48 & SG-HVS48 &  241.981170 & $-$14.429 & $-$32.566 & $-$282.39 & 3.72 & 6273 & 3.74 & $-$1.86 & $+$0.13 & $-$105.27 & 516.44 & this work \\
& 1403645069530607616 & 51.582219 & 0.063 & 0.082 & 3.10 & 0.76 & 20 & 0.14 & 0.05 & 0.03 & 3.10 & 122.29 & \\
\multirow{1}{*}
49 & SG-HVS49 &  252.688090 & $-$6.120 & $-$38.142 & $-$126.76 & 3.37 & 5938 & 4.51 & $-$1.87 & $+$0.27 & 55.59 & 526.38 & this work \\
& 1356298174692911104 & 41.324670 & 0.056 & 0.066 & 2.77 & 0.71 & 14 & 0.02 & 0.00 & 0.01 & 2.77 & 124.49 & \\
\multirow{1}{*}
50 & SG-HVS50 &  319.051060 & 2.739 & $-$66.662 & $-$303.65 & 2.29 & 5407 & 4.24 & $-$2.00 & $+$0.41 & $-$133.28 & 555.37 & this work \\
& 2689713751472944384 & 0.499367 & 0.104 & 0.086 & 4.09 & 0.55 & 63 & 0.11 & 0.05 & 0.02 & 4.09 & 169.17 & \\
\multirow{1}{*}
51 & SG-HVS51 &  338.674800 & $-$17.371 & $-$48.307 & $-$167.53 & 3.00 & 6201 & 2.06 & $-$1.24 & $+$0.21 & $-$42.54 & 574.38 & this work \\
& 2609260664602088704 & $-$8.696933 & 0.056 & 0.058 & 2.84 & 0.53 & 58 & 0.69 & 0.03 & 0.01 & 2.84 & 124.62 & \\
\multirow{1}{*}
52 & SG-HVS52 &  341.939260 & 36.388 & $-$5.165 & $-$91.33 & 3.36 & 5913 & 3.44 & $-$2.08 & $+$0.18 & 134.75 & 520.55 & this work$^a$ \\
 & 1888115422016149248 & 31.152207 & 0.020 & 0.025 & 2.79 & 0.29 & 33 & 0.07 & 0.05 & 0.01 & 2.79 & 48.67 & \\
 \cline{1-14}
\multicolumn{14}{c}{HVSs or candidates from other literature} \\
\cline{1-14}
\multirow{1}{*}
53 & J013655.91+242546.0 &  24.232958 & $-$1.820 & $-$6.660 & 324.00 & 10.90 & 9100 & 3.90 & -- & -- & 456.42 & 557.94 & \citet{Tillich2009} \\
& 291821209329550464 & 24.429440 & 0.050 & 0.041 & 5.90 & 2.00 & 250 & 0.15 & -- & -- & 5.90 & 30.35 & \\
\multirow{1}{*}
54 & LAMOST-HVS3 &  50.321157 & 1.146 & $-$0.549 & 361.00 & 22.32 & 14000 & -- & -- & -- & 407.96 & 428.93 & \citet{Huang2017} \\
& 56282900715073664 & 19.126719 & 0.082 & 0.067 & 12.52 & 2.50 & -- & -- & -- & -- & 12.52 & 12.23 & \\
\multirow{1}{*}
55 & HVS3(HE 0437-5439) &  69.553330 & 0.853 & 1.614 & 723.00 & 61.00 & 20354 & 3.77 & -- & -- & 529.95 & 747.09 & \citet{Edelmann2005} \\
& 4777328613382967040 & $-$54.553300 & 0.049 & 0.061 & 3.00 & 12.00 & 360 & 0.05 & -- & -- & 3.00 & 69.17 & \\
\multirow{1}{*}
56 & HD 271791 &  90.616163 & $-$0.413 & 4.704 & 441.00 & 24.00 & 17810 & 3.04 & -- & -- & 221.41 & 569.75 & \citet{Heber2008} \\
& 5284151216932205312 & $-$66.791300 & 0.035 & 0.042 & -- & 2.50 & 180 & 0.03 & -- & -- & -- & 50.48 & \\
\multirow{1}{*}
57 & HVS1 &  136.937400 & $-$0.604 & $-$0.474 & 833.00 & 101.30 & 11125 & 3.91 & -- & -- & 673.13 & 731.61 & \citet{Brown2005} \\
& 577294697514301440 & 2.751950 & 0.602 & 0.385 & 5.50 & 14.46 & 463 & 0.20 & -- & -- & 5.50 & 109.15 & \\
\multirow{1}{*}
58 & LAMOST-HVS1 &  138.027170 & $-$3.557 & $-$0.793 & 620.00 & 13.40 & 20700 & 3.67 & $-$0.13 & -- & 483.45 & 554.21 & \citet{Zheng2014} \\
& 590511484409775360 & 9.272722 & 0.031 & 0.027 & 10.00 & 2.20 & 1200 & 0.19 & 0.07 & -- & 10.00 & 14.68 & \\
\multirow{1}{*}
59 & HVS4 &  138.254200 & $-$0.204 & $-$0.601 & 600.90 & 63.80 & 14547 & 4.15 & -- & -- & 551.94 & 558.30 & \citet{Brown2006} \\
& 699811079173836928 & 30.855500 & 0.263 & 0.193 & 6.20 & 9.99 & 607 & 0.21 & -- & -- & 6.20 & 13.43 & \\
\multirow{1}{*}
60 & HVS5 &  139.497800 & 0.001 & $-$0.989 & 545.50 & 44.40 & 12000 & 3.89 & -- & -- & 653.45 & 653.87 & \citet{Brown2006} \\
& 1069326945513133952 & 67.377300 & 0.084 & 0.105 & 4.30 & 5.11 & 350 & 0.13 & -- & -- & 4.30 & 4.48 & \\
\multirow{1}{*}
61 & HVS2(US708) &  143.336958 & $-$5.447 & 1.776 & 917.00 & 8.50 & 47200 & 5.69 & -- & -- & 927.88 & 1001.57 & \citet{Geier2015} \\
& 815106177700219392 & 44.284861 & 0.193 & 0.164 & 7.00 & 1.00 & 400 & 0.09 & -- & -- & 7.00 & 10.10 & \\
\multirow{1}{*}
62 & HVS8 &  145.558500 & $-$0.877 & $-$0.276 & 499.30 & 53.43 & 11000 & 3.75 & -- & -- & 407.24 & 478.67 & \citet{Brown2007} \\
& 633599760258827776 & 20.056100 & 0.162 & 0.144 & 2.90 & 9.84 & 1000 & 0.25 & -- & -- & 2.90 & 23.53 & \\
\multirow{1}{*}
63 & HVS9 &  155.404500 & 0.265 & $-$0.808 & 616.80 & 75.13 & 11680 & 3.83 & -- & -- & 455.27 & 491.18 & \citet{Brown2007} \\
& 3830584196322129920 & $-$0.876330 & 0.427 & 0.648 & 5.10 & 11.76 & 529 & 0.21 & -- & -- & 5.10 & 81.39 & \\
\multirow{1}{*}
64 & HVS21 &  158.576042 & $-$0.199 & $-$0.652 & 356.80 & 108.64 & 13229 & 4.16 & -- & -- & 392.90 & 405.83 & \citet{Brown2012} \\
& 834069905715968640 & 48.192936 & 0.412 & 0.650 & 7.50 & 21.01 & 998 & 0.31 & -- & -- & 7.50 & 127.82 & \\
\multirow{1}{*}
65 & HVS14 &  161.007300 & $-$2.166 & 2.282 & 537.30 & 102.19 & 11030 & 3.90 & -- & -- & 406.18 & 1681.51 & \citet{Brown2009} \\
& 3859275333773935488 & 6.194167 & 1.376 & 1.684 & 7.20 & 16.47 & 554 & 0.24 & -- & -- & 7.20 & 719.69 & \\
\multirow{1}{*}
66 & HVS12 &  162.540000 & 0.928 & $-$0.193 & 552.20 & 64.54 & 12098 & 4.62 & -- & -- & 412.92 & 565.64 & \citet{Brown2009} \\
& 3809777626689513216 & 3.264080 & 0.876 & 0.580 & 6.60 & 8.32 & 632 & 0.28 & -- & -- & 6.60 & 175.84 & \\
\multirow{1}{*}
67 & HVS13 &  163.201300 & 0.065 & $-$0.203 & 572.70 & 104.14 & 11241 & 4.04 & -- & -- & 423.34 & 446.36 & \citet{Brown2009} \\
& 3804790100211231104 & $-$0.026090 & 0.788 & 0.629 & 4.50 & 18.70 & 739 & 0.33 & -- & -- & 4.50 & 177.00 & \\
\multirow{1}{*}
68 & HVS6 &  166.489400 & 0.119 & 0.125 & 609.40 & 55.36 & 12190 & 4.30 & -- & -- & 498.10 & 566.23 & \citet{Brown2006} \\
& 3867267443277880320 & 9.577640 & 0.298 & 0.231 & 6.80 & 7.14 & 546 & 0.23 & -- & -- & 6.80 & 31.84 & \\
\multirow{1}{*}
69 & HVS24 &  167.901830 & 0.101 & $-$0.403 & 496.20 & 61.80 & 11103 & 4.00 & -- & -- & 357.17 & 383.87 & \citet{Brown2014} \\
& 3810351984075984768 & 0.982333 & 0.311 & 0.265 & 6.80 & 11.10 & 806 & 0.31 & -- & -- & 6.80 & 33.78 & \\
\multirow{1}{*}
70 & HVS7 &  173.300500 & $-$0.089 & 0.020 & 527.80 & 52.17 & 12000 & 3.80 & -- & -- & 398.54 & 451.19 & \citet{Brown2006} \\
& 3799146650623432704 & 1.140250 & 0.183 & 0.129 & 2.70 & 6.49 & 500 & 0.10 & -- & -- & 2.70 & 15.26 & \\
\multirow{1}{*}
71 & HVS15 &  173.421200 & $-$1.295 & $-$0.483 & 461.00 & 61.00 & 11132 & 4.05 & -- & -- & 323.82 & 423.76 & \citet{Brown2009} \\
& 3794074603484360704 & $-$1.353940 & 0.357 & 0.232 & 3.00 & 7.00 & 535 & 0.23 & -- & -- & 3.00 & 70.93 & \\
\multirow{1}{*}
72 & HVS19 &  173.823958 & 0.517 & $-$0.980 & 592.80 & 96.87 & 12900 & 4.53 & -- & -- & 488.21 & 645.50 & \citet{Brown2012} \\
& 3911105521632982400 & 8.033747 & 1.079 & 1.140 & 11.80 & 15.17 & 793 & 0.29 & -- & -- & 11.80 & 317.50 & \\
\multirow{1}{*}
73 & HVS20 &  174.154708 & $-$0.183 & $-$0.990 & 512.10 & 75.40 & 11149 & 4.21 & -- & -- & 392.39 & 425.34 & \citet{Brown2012} \\
& 3800802102817768832 & 3.518567 & 0.656 & 0.562 & 8.50 & 11.11 & 649 & 0.28 & -- & -- & 8.50 & 110.68 & \\
\multirow{1}{*}
74 & HVS22 &  175.443542 & 0.065 & $-$0.590 & 597.80 & 83.98 & 11145 & 4.35 & -- & -- & 484.57 & 504.51 & \citet{Brown2014} \\
& 3897063727354575488 & 4.704803 & 0.875 & 0.674 & 13.40 & 13.54 & 859 & 0.30 & -- & -- & 13.40 & 139.37 & \\
\multirow{1}{*}
75 & HVS10 &  180.907700 & $-$1.086 & $-$0.991 & 467.90 & 51.76 & 11270 & 4.38 & -- & -- & 413.70 & 439.40 & \citet{Brown2007} \\
& 3926757653770374272 & 18.047330 & 0.451 & 0.207 & 5.60 & 5.72 & 533 & 0.23 & -- & -- & 5.60 & 36.45 & \\
\cline{1-14}
\end{tabular}}
\end{threeparttable}
\end{table*}

\begin{table*}[!b]
\centering
\begin{threeparttable}
\tablename{ A1: Basic Parameters for 88 HVSs or candidates.}
\label{tbl:table2}
\resizebox{\textwidth}{!}{
\begin{tabular}{lrrrrrrrrrrrrr}
\toprule
\multirow{1}{*} ID & Notation & $\rm{ra}$ & $\mu_{\alpha}\rm{cos}\,\delta$ & $\mu_{\delta}$ & $v_{\rm{los}}$ & $\it{d}$  & $T_{\rm eff}$ & ${\rm log} g$ & [Fe/H]  & [$\alpha$/Fe]& $\it{V}_{\rm{rf}}$ & $\it{V}_{\rm{GSR}}$ & Reference\\ 
&  Gaia ID & $\rm{dec}$ & $\sigma_{\mu_{\alpha}\rm{cos}\,\delta}$ & $\sigma_{\mu_{\delta}}$  & $\sigma_{v_{\rm{los}}}$ & $\sigma_{\it{d}}$ & $\sigma_{T_{\rm eff}}$ & $\sigma_{{\rm log} g}$ & $\sigma_{\rm{[Fe/H]}}$  & $\sigma_{[\alpha/\rm{Fe}]}$ & $\sigma_{\it{V}_{\rm{rf}}}$& $\sigma_{\it{V}_{\rm{GSR}}}$ &  \\
& & $(\rm{deg})$ & $(\rm{mas\,yr^{-1}})$ & $(\rm{mas\,yr^{-1}})$ & $(\rm{km\,s^{-1}})$ & $(\rm{kpc})$ & (K) & (dex) & (dex)  & (dex)& $(\rm{km\,s^{-1}})$ & $(\rm{km\,s^{-1}})$ &  \\
\cline{1-14}
\multirow{1}{*}
76 & HIP 60350 &  185.623360 & $-$13.304 & 15.033 & 262.00 & 3.10 & 16100 & 4.10 & -- & -- & 302.95 & 532.43 & \citet{Irrgang2010} \\
& 1533367925276710272 & 40.826545 & 0.041 & 0.047 & 5.00 & 0.60 & 500 & 0.15 & -- & -- & 5.00 & 39.86 & \\
\multirow{1}{*}
77 & HVS16 &  186.347500 & $-$1.290 & $-$0.535 & 429.80 & 65.00 & 10388 & 3.96 & -- & -- & 341.81 & 434.07 & \citet{Brown2009} \\
& 3708104343359742848 & 5.376056 & 0.501 & 0.294 & 7.00 & 6.00 & 666 & 0.29 & -- & -- & 7.00 & 97.26 & \\
\multirow{1}{*}
78 & LP40-365 &  211.647710 & $-$49.569 & 148.642 & 498.00 & 0.30 & -- & -- & -- & -- & 667.58 & 737.40 & \citet{Vennes2017} \\
& 1711956376295435520 & 74.316110 & 0.029 & 0.029 & 1.10 & 0.11 & -- & -- & -- & -- & 1.10 & 28.72 & \\
\multirow{1}{*}
79 & HVS23 &  240.537580 & -- & -- & 259.30 & 114.87 & 10996 & 3.99 & -- & -- & 307.14 & -- & \citet{Brown2014} \\
& -- & 0.912272 & -- & -- & 9.80 & 20.10 & 778 & 0.29 & -- & -- & 9.80 & -- & \\
\multirow{1}{*}
80 & J1603-6613 &  240.766917 & 39.872 & $-$7.176 & $-$485.00 & 1.77 & 10590 & 5.34 & -- & -- & $-$637.99 & 811.01 & \citet{Raddi2019} \\
& 5822236741381879040 & $-$66.224139 & 0.048 & 0.076 & 5.00 & 0.34 & 370 & 0.20 & -- & -- & 5.00 & 38.24 & \\
\multirow{1}{*}
81 & LAMOST-HVS2 &  245.086520 & $-$2.398 & $-$0.803 & 341.00 & 22.24 & 20600 & -- & -- & -- & 501.31 & 510.77 & \citet{Huang2017} \\
& 1330715287893559936 & 37.794456 & 0.027 & 0.033 & 7.79 & 4.57 & -- & -- & -- & -- & 7.79 & 10.93 & \\
\multirow{1}{*}
82 & D6-1 &  249.381980 & $-$80.232 & $-$195.960 & 1200.00 & 1.87 & -- & -- & -- & -- & 1038.44 & 2001.47 & \citet{Shen2018} \\
& 5805243926609660032 & $-$74.343490 & 0.060 & 0.064 & 40.00 & 0.25 & -- & -- & -- & -- & 40.00 & 219.52 & \\
\multirow{1}{*}
83 & HVS17 &  250.484958 & $-$1.129 & $-$0.930 & 250.20 & 49.59 & 12350 & 3.80 & -- & -- & 437.00 & 474.02 & \citet{Brown2012} \\
& 1407293627068696192 & 47.396140 & 0.090 & 0.096 & 2.90 & 4.34 & 290 & 0.09 & -- & -- & 2.90 & 14.37 & \\
\multirow{1}{*}
84 & D6-3 &  283.007850 & 9.406 & 211.788 & $-$20.00 & 2.52 & -- & -- & -- & -- & 212.53 & 2510.79 & \citet{Shen2018} \\
& 2156908318076164224 & 62.036168 & 0.136 & 0.149 & 80.00 & 0.65 & -- & -- & -- & -- & 80.00 & 660.07 & \\
\multirow{1}{*}
85 & D6-2 &  324.612490 & 98.285 & 240.182 & 20.00 & 0.83 & -- & -- & -- & -- & 250.98 & 1127.32 & \citet{Shen2018} \\
& 1798008584396457088 & 25.373712 & 0.068 & 0.058 & 60.00 & 0.04 & -- & -- & -- & -- & 60.00 & 52.77 & \\
\multirow{1}{*}
86 & S5-HVS1 &  343.715345 & 35.406 & 0.535 & 1017.00 & 8.88 & 9630 & 4.23 & 0.29 & -- & 964.19 & 1727.36 & \citet{Koposov2020} \\
& 6513109241989477504 & $-$51.195610 & 0.029 & 0.038 & 2.70 & 0.01 & 110 & 0.03 & 0.08 & -- & 2.70 & 2.32 & \\
\multirow{1}{*}
87 & LAMOST-HVS4 &  344.656500 & 0.133 & $-$0.338 & 359.00 & 27.90 & 15140 & 3.90 & 0.29 & -- & 592.12 & 595.77 & \citet{Li2018} \\
& 1928660566125735680 & 40.001470 & 0.057 & 0.063 & 7.00 & 1.50 & 578 & 0.30 & 0.18 & -- & 7.00 & 7.06 & \\
\multirow{1}{*}
88 & HVS18 &  352.270583 & 0.007 & $-$0.239 & 237.30 & 77.70 & 11993 & 4.08 & -- & -- & 452.39 & 463.23 & \citet{Brown2012} \\
& 2872564390598678016 & 33.003186 & 0.337 & 0.316 & 6.40 & 11.09 & 516 & 0.22 & -- & -- & 6.40 & 33.84 & \\
\cline{1-14}
\end{tabular}}
\begin{tablenotes}\footnotesize
\item[$^a$] This candidate is also reported by Li21.
\item[$^b$] This candidate is also reported by \citet{Bromley2018} and Li21.
\end{tablenotes}
\end{threeparttable}
\end{table*}

\end{appendix}

\end{document}